\documentclass[12pt]{article}
\pdfoutput=1
\usepackage{makeidx}
\makeindex
\usepackage[a4paper]{geometry}
\usepackage{jheppub,amsmath,amssymb,amsfonts,amsxtra,mathrsfs,graphics,graphicx,amsthm,epsfig,bm,longtable,float,color,tikz,mathtools,xfrac,footnote,rotating,lscape}
\usepackage{dsfont}
\usetikzlibrary{decorations.pathmorphing}
\usetikzlibrary{decorations.markings}
\usetikzlibrary{arrows, decorations.markings, calc, fadings, decorations.pathreplacing, patterns, decorations.pathmorphing, positioning}
\usepackage{tikz-cd}

\usepackage{fixmath} 
\usepackage{scalerel}
\newlength\bshft
\def\fakebold#1{\ThisStyle{\ooalign{$\SavedStyle#1$\cr%
  \kern-\bshft$\SavedStyle#1$\cr%
  \kern\bshft$\SavedStyle#1$}}}

\usetikzlibrary{positioning,shapes}
\usetikzlibrary{chains}
\usetikzlibrary{arrows,fit,decorations.pathreplacing}
\tikzstyle{every picture}+=[remember picture]
\tikzstyle{na} = [baseline=-.5ex]

\usepackage{empheq}
\usepackage{multirow}
\usepackage{booktabs}
\usepackage[american]{babel}

\usepackage[latin1]{inputenc}

\makeatletter
\newcommand{\vast}{\bBigg@{1}}
\newcommand{\Vast}{\bBigg@{5}}
\makeatother

\usepackage{hyperref}
\renewcommand{\arraystretch}{1.2}
\numberwithin{equation}{section}

\newcommand{\ie}{\textit{i.e.}}

\numberwithin{equation}{section}

\newcommand{\be}{\begin{equation}} \newcommand{\ee}{\end{equation}}
\newcommand{\bea}{\begin{equation} \begin{aligned}} \newcommand{\eea}{\end{aligned} \end{equation}}

\def\U{\mathrm{U}}

\def\SU{\mathrm{SU}}

\def\USp{\mathrm{USp}}

\newcommand{\vol}{\mathrm{vol}}

\newcommand{\wt}{\widetilde}

\DeclareMathOperator{\Tr}{Tr}

\DeclareMathOperator{\sign}{sign}

\DeclareMathOperator{\Li}{Li}

\newcommand{\cA}{\mathcal{A}}
\newcommand{\cB}{\mathcal{B}}
\newcommand{\cC}{\mathcal{C}}

\newcommand{\cF}{\mathcal{F}}
\newcommand{\cG}{\mathcal{G}}
\newcommand{\cH}{\mathcal{H}}
\newcommand{\cI}{\mathcal{I}}

\newcommand{\cM}{\mathcal{M}}
\newcommand{\cN}{\mathcal{N}}

\newcommand{\cQ}{\mathcal{Q}}

\newcommand{\cS}{\mathcal{S}}

\newcommand{\cW}{\mathcal{W}}
\newcommand{\cX}{\mathcal{X}}

\newcommand{\cZ}{\mathcal{Z}}

\newcommand{\fg}{\mathfrak{g}}

\newcommand{\fh}{\mathfrak{h}}

\newcommand{\fm}{\mathfrak{m}}

\newcommand{\fn}{\mathfrak{n}}

\newcommand{\fp}{\mathfrak{p}}

\newcommand{\fR}{\mathfrak{R}}
\newcommand{\fs}{\mathfrak{s}}
\newcommand{\ft}{\mathfrak{t}}

\newcommand{\ii}{\mathrm{i}}
\newcommand{\diff}{\mathrm{d}}
\usepackage[Symbol]{upgreek}
\usepackage{bm}

\usepackage{subfigure}

\usepackage{calligra}
\DeclareMathAlphabet{\mathcalligra}{T1}{calligra}{m}{n}

\theoremstyle{plain}

  \theoremstyle{definition}

\providecommand{\examplename}{Example}
\providecommand{\theoremname}{Theorem}

\makeatletter
\g@addto@macro\bfseries{\boldmath}
\makeatother

\title{Black hole microstate counting in Type IIB from 5d SCFTs}
\author[a]{Martin Fluder,}
\author[a]{Seyed Morteza Hosseini}
\author[b]{and Christoph F.~Uhlemann}
\affiliation[a]{Kavli IPMU (WPI), UTIAS, The University of Tokyo,\\Kashiwa, Chiba 277-8583, Japan}
\affiliation[b]{Mani L. Bhaumik Institute for Theoretical Physics,\\Department of Physics and Astronomy,\\University of California, Los Angeles, CA 90095, USA}
\emailAdd{martin.fluder@ipmu.jp}
\emailAdd{morteza.hosseini@ipmu.jp}
\emailAdd{uhlemann@physics.ucla.ed}

\preprint{IPMU19-0020}

\abstract{
We use recently established AdS$_6$/CFT$_5$ dualities to count the microstates of magnetically charged AdS$_6 \times S^2 \times \Sigma$ black holes in Type IIB. The near-horizon limit is described by solutions with AdS$_2 \times \Sigma_{\mathfrak{g}_1} \times \Sigma_{\mathfrak{g}_2} \times S^2 \times \Sigma$ geometry, where $\Sigma_{\mathfrak{g}_i}$ are Riemann surfaces of constant curvature and $\Sigma$ is a further Riemann surface over which the geometry is warped. 
Our results show that the topologically twisted indices of the proposed dual superconformal field theories precisely reproduce the Bekenstein-Hawking entropy of this class of black holes. 
This provides further support for a prescription to compute five-dimensional topologically twisted indices put forth recently, and for the proposed dualities. We confirm the $N^4$ scaling found in the sphere partition functions and extend previous matches of sphere partition functions to AdS$_6$ solutions with monodromy.
}

\begin{document}

\setcounter{tocdepth}{2}
\maketitle

%
%

\date{Dated: \today}

\section{Introduction}
\label{sect:intro}

The study of exactly computable quantities in supersymmetric field theories has led to dramatic progress in the non-perturbative understanding of quantum field theory. Partition functions on compact Euclidean spaces and supersymmetric indices are among the best understood examples, in particular in conjunction with AdS/CFT dualities, where they allow for rigorous tests of string theory constructions and holographic correspondences.

Our focus in this work is on sphere partition functions and topologically twisted indices of five-dimensional superconformal field theories (SCFTs).
The topologically twisted index~\cite{Benini:2015noa,Benini:2016hjo} of the ABJM theory~\cite{Aharony:2008ug} has been shown to count the microstates of magnetically charged AdS$_4$ black holes~\cite{Cacciatori:2009iz}
in the holographically dual gravitational theory in~\cite{Benini:2015eyy,Benini:2016rke}.
This was soon generalized to other three- and four-dimensional gauge theories~\cite{Hosseini:2016tor,Hosseini:2016ume, Hosseini:2016cyf,Hosseini:2018qsx,Hong:2018viz,Cabo-Bizet:2017jsl,Azzurli:2017kxo,Hosseini:2017fjo,Benini:2017oxt,Bobev:2018uxk,Toldo:2017qsh,Gang:2018hjd},%
\footnote{For other interesting developments in this context see~\cite{Liu:2017vbl,Liu:2017vll,Jeon:2017aif, Halmagyi:2017hmw,Cabo-Bizet:2017xdr, Hristov:2018lod,Nian:2017hac,Liu:2018bac}.} as well as the five-dimensional $\USp(2N)$ (``Seiberg") theories which can be realized by D4-D8-O8 systems and have holographic duals in massive type IIA supergravity~\cite{Hosseini:2018uzp,Suh:2018tul,Hosseini:2018usu}. Specifically, the topologically twisted index of $\cN=1$ gauge theories in five dimensions is given by the partition function on $\cM_4 \times S^1$ with a partial topological twist on $\cM_4$.
The manifold $\cM_4$ is either toric K\"ahler~\cite{Hosseini:2018uzp}, or a product of two Riemann surfaces~\cite{Hosseini:2018uzp,Crichigno:2018adf}.
The twisted index is a function of background magnetic fluxes and chemical potentials for the R- and global symmetries of the theory.
For five-dimensional $\USp(2N)$ theories, it is expected that the twisted index accounts for the entropy of a class of magnetic AdS$_6$ black holes in massive type IIA supergravity with AdS$_2\times \cM_4$ near-horizon region. Using the consistent truncation from Type IIA to Romans' six-dimensional $F(4)$ gauged supergravity~\cite{Romans:1985tw}, this was indeed confirmed in~\cite{Suh:2018tul,Hosseini:2018uzp,Hosseini:2018usu} (see also~\cite{Suh:2018qyv}). Within six-dimensional $F(4)$ gauged supergravity, there is a universal relation between the black hole entropy and the on-shell action which holographically computes the five-sphere partition function~\cite{Suh:2018tul} (see~\eqref{universal:logZ:FS5}).%
\footnote{Similar universal relations were discussed in a variety of dimensions in~\cite{Bobev:2017uzs}. However, the AdS$_2$ solutions to six-dimensional gauged supergravity used there turned out to be incomplete, leading to an incorrect expression for the Bekenstein-Hawking entropy in this particular case.}
This relation confirms the field theory prediction for the (leading order) large $N$ behavior of the topologically twisted index and the five-sphere partition function~\cite{Hosseini:2018uzp}.

A more general construction of five-dimensional SCFTs is via 5-brane webs in Type IIB~\cite{Aharony:1997ju,Kol:1997fv,Aharony:1997bh} (alternative constructions are based in M-theory~\cite{Intriligator:1997pq,Brandhuber:1997ua,Jefferson:2017ahm,Jefferson:2018irk,Xie:2017pfl,Apruzzi:2018nre,Closset:2018bjz}). Infinite families of five-dimensional SCFTs can be engineered via 5-brane webs, and the construction may be further generalized by including 7-branes~\cite{DeWolfe:1999hj}. Corresponding to this large class of field theories, one expects large classes of AdS$_6$ solutions and corresponding AdS$_6$/CFT$_5$ dualities in Type IIB. Such classes of AdS$_6$ solutions have been constructed in~\cite{DHoker:2016ujz,DHoker:2016ysh,DHoker:2017mds,DHoker:2017zwj},\footnote{Earlier analyses of the BPS equations can be found in~\cite{Apruzzi:2014qva,Kim:2015hya,Kim:2016rhs}. T-duals of the Type IIA solution~\cite{Brandhuber:1999np} have been discussed in~\cite{Lozano:2012au,Lozano:2013oma}.} and permit a precise identification with 5-brane constructions. Various aspects of these dualities have since been studied~\cite{Gutperle:2017tjo,Kaidi:2017bmd,Apruzzi:2018cvq,Gutperle:2018vdd,Gutperle:2018wuk,Kaidi:2018zkx,Lozano:2018pcp,Chen:2019one}, and they have been subjected to explicit quantitative tests in~\cite{Bergman:2018hin,Fluder:2018chf}. The supergravity solutions have geometries of the form ${\rm AdS}_6\times S^2$ warped over a Riemann surface $\Sigma$, which encodes the structure of the associated 5-brane web. Consistent truncations to Romans' six-dimensional $F(4)$ gauged supergravity for arbitrary such solutions were constructed in~\cite{Hong:2018amk,Malek:2018zcz}. Truncations with additional vector multiplets were discussed recently in~\cite{Malek:2019ucd}. The existence of these truncations implies that AdS/CFT predicts the same universal relation between the  five-sphere partition function and the topologically twisted index that was found for the five-dimensional Seiberg theories to also hold for five-dimensional SCFTs with holographic duals in type IIB supergravity.

In this paper, we verify this prediction from the field theory side. We select a representative sample of five-dimensional SCFTs engineered in Type IIB and show that for $\cM_4$ the product of two constant curvature Riemann surfaces, $\cM_4=\Sigma_{\fg_1} \times \Sigma_{\fg_2}$, the large $N$ relation between the topologically twisted index and the five-sphere partition function,
\be
 \label{universal:logZ:FS5}
 \log Z_{\Sigma_{\fg_1} \times \Sigma_{\fg_2} \times S^1} \ = \ - \frac{8}{9} ( 1- \fg_1 ) ( 1 - \fg_2 ) F_{S^5} \, ,
\ee
holds for this class of theories as well. Our sample of theories includes the (unconstrained) $T_N$ theories~\cite{Benini:2009gi}, and the theories realized on intersections of D5 and NS5 branes~\cite{Aharony:1997bh} as examples that are realized by 5-branes only. That the S$^5$ partition functions of these theories match the prediction of the putative holographic duals was shown in~\cite{Fluder:2018chf}, and here we explicitly confirm that the topologically twisted index matches as well. Furthermore, we include a class of constrained $T_N$ theories, that we refer to as $T_{N,K,j}$, which are engineered using 5-branes and 7-branes in Type IIB and whose holographic duals are solutions with monodromy~\cite{Chaney:2018gjc}. For these theories we compute the $S^5$ partition functions, the topologically twisted indices and the corresponding quantities in supergravity, and show that they match.

Our results provide further examples where the topologically twisted index counts the microstates of black holes in the dual gravitational theory. Let us emphasize, that our computation of the topologically twisted index as the partition function on $\Sigma_{\fg_1} \times \Sigma_{\fg_2} \times S^1$ is based on the Bethe ansatz equations involving the effective Seiberg-Witten prepotential of the four-dimensional theory resulting from the compactification on $S^1$ as proposed in \cite{Hosseini:2018uzp}.\footnote{We refer the reader to section~\ref{bethesum} for more details.} The match to the Bekenstein-Hawking entropy  provides further support for the constructions put forward in \cite{Hosseini:2018uzp}, and for the AdS$_6$/CFT$_5$ dualities of~\cite{DHoker:2016ujz,DHoker:2017mds,DHoker:2017zwj,Chaney:2018gjc}. The (by now) large class of field theories for which \eqref{universal:logZ:FS5} holds at large $N$ suggests that this relation may be completely universal at large $N$ for five-dimensional SCFTs. Additionally, our results for the $T_{N,K,j}$ theories extend previous matches of the sphere partition functions to AdS$_6$ solutions with monodromy.

The remainder of the paper is organized as follows. 
In section~\ref{sec:fieldtheory} we review the five-dimensional SCFTs under consideration and discuss their gauge theory deformations
and the computation of the topologically twisted indices, using matrix model techniques.
In section~\ref{sec:sugra} we discuss magnetic AdS$_6$ black holes in Type IIB and their Bekenstein-Hawking entropies.
For the  $T_{N,K,j}$ theories we holographically compute the $S^5$ partition function.
We close with a discussion in section~\ref{sec:discussion}. The explicit expressions for the matrix models can be found in the appendix.

\section{Topologically twisted indices of 5d SCFTs}\label{sec:fieldtheory}

In this section, we discuss three examples of five-dimensional SCFTs engineered from $(p,q)$ 5-brane junctions and compute the topologically twisted indices at large $N$. The examples include the intersection of $N$ D5-branes with $M$ NS5-branes discussed initially in~\cite{Aharony:1997bh}, which we refer to as $\#_{N,M}$. We also include the unconstrained $T_N$ theories of~\cite{Benini:2009gi}, and a subset of the theories that can be obtained by Higgs-branch flows from the $T_N$ theories, which we will refer to as $T_{N,K,j}$~\cite{Chaney:2018gjc}. The $\#_{N,M}$ theories include the rank-$1$ $E_5$ theory, the $T_N$ theories include the rank-$1$ $E_6$ theory and the $T_{N,K,j}$ theories include the rank-$1$ $E_7$ theory.

\subsection{The \texorpdfstring{$\#_{N,M}$}{\#[N,M]}, \texorpdfstring{$T_{N}$}{T[N]} and \texorpdfstring{$T_{N,K,j}$}{T[N,K,j]} theories}\label{sec:quivers}

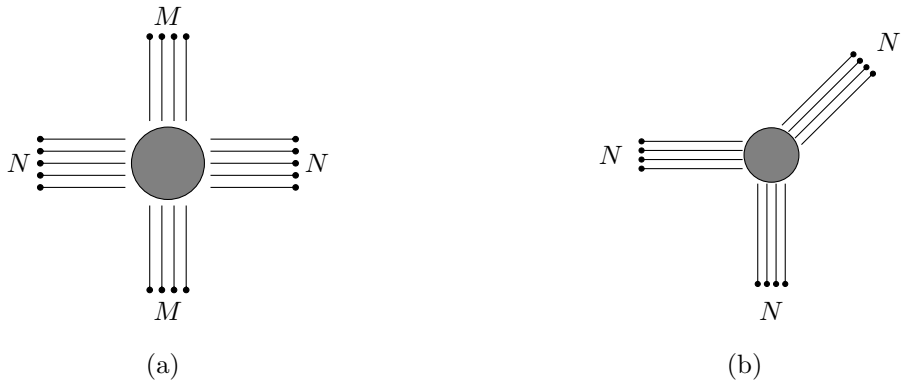
\begin{figure}
\centering
\subfigure[][]{\label{fig:D5NS5-junction}
  \begin{tikzpicture}[scale=0.8]
  \draw[fill=gray] (0,0) circle (0.6);
  \foreach \i in {-5,-5/2,0,5/2,5}{
   \draw (0.7,0-0.08*\i) -- +(1.4,0) [fill] circle (1.3pt);
   \draw (-0.7,0-0.08*\i) -- +(-1.4,0) [fill] circle (1.3pt);
  }
  \foreach \i in {-5,-1.667,1.667,5}{
   \draw (0.06*\i,0.7) -- +(0,1.4) [fill] circle (1.3pt);
   \draw (0.06*\i,-0.7) -- +(0,-1.4) [fill] circle (1.3pt);
  }
  \node at (0.7+1.75,0) {\footnotesize $N$};
  \node at (-0.7-1.75,0) {\footnotesize $N$};
  \node at (0,0.7+1.75) {\footnotesize $M$};
  \node at (0,-0.7-1.75) {\footnotesize $M$};
  \end{tikzpicture}
}\hskip 30mm
\subfigure[][]{\label{fig:TN-junction}
\begin{tikzpicture}[scale=0.95]
    \foreach \i in {-3/2,-1/2,1/2,3/2}{
      \draw (-0.4,0.127*\i) -- (-1.8,0.127*\i) [fill=black] circle (1pt);
      \draw (0.127*\i,-0.4) -- (0.127*\i,-1.8) [fill=black] circle (1pt);
      \draw (0.28+0.09*\i,0.28-0.09*\i) -- (1.27+0.09*\i,1.27-0.09*\i) [fill=black] circle (1pt) ;
    }
    \draw[fill=gray] (0,0) circle (0.38);
    
    \node [anchor=east] at (-1.9,0) {\footnotesize $N$};
    \node [anchor=south west] at (1.3,1.3) {\footnotesize $N$};
    \node [anchor=north] at (0,-1.9) {\footnotesize $N$};
\end{tikzpicture}
}
\caption{5-brane junctions: (a) $\#_{N,M}$ theory and (b) unconstrained $T_N$ theories.}
\end{figure}

\paragraph*{$\#_{N,M}$ theory:} The 5-brane junction realizing the $\#_{N,M}$ theory is shown in figure~\ref{fig:D5NS5-junction}. 
The SCFT in general has global symmetry $\SU(M)^2\times \SU(N)^2\times \U(1)$, which may be further enhanced for small values of $N$ and $M$. A relevant deformation flowing to a gauge theory in the infrared yields the linear quiver
\begin{align}\label{eq:D5NS5-quiver}
[N]\stackrel{y_1}{-}(N)\stackrel{x_1}{-}\cdots \stackrel{x_{M-2}}{-}(N)\stackrel{y_2}{-}[N] \,.
\end{align}
Here and in the following $(N)$ denotes an $\SU(N)$ gauge node and $[K]$ denotes $K$ hypermultiplets in the fundamental representation of the gauge group node they are attached to. 
There are bi-fundamental hypermultiplets between adjacent gauge group nodes which we denote by $x_i$, and by $y_i$ we denote the fundamental hypermultiplets.
The quiver in \eqref{eq:D5NS5-quiver} has a total of $M-1$ $\SU(N)$ gauge nodes and the Chern-Simons levels are zero for all nodes.
For $N=M=2$, this is the $\SU(2)$ Seiberg theory with global symmetry enhanced to $E_5$.

\paragraph*{$T_{N}$ theory:} The five-dimensional $T_N$ theories are realized on an intersection of $N$ D5-branes, $N$ NS5-branes and $N$ $(1,1)$ 5-branes, see figure~\ref{fig:TN-junction}. Upon $\rm S^1$ compactification they reduce to the well-known four-dimensional $T_N$ theories~\cite{Gaiotto:2009we}, which can be constructed by compactifying the six-dimensional $\cN=(2,0)$ theory on a three-punctured sphere.
The types of the punctures are encoded in the five-dimensional theories in the combinatorics of how the 5-branes are terminated on 7-branes. We first discuss the case where each 5-brane is terminated on an individual 7-brane. In that case there are no constraints from the $s$-rule \cite{Benini:2009gi}, which imposes for example that at most one D5-brane can stretch between an NS5-brane and a D7-brane. The global symmetry is at least $\SU(N)^3$, and we will refer to this theory simply as $T_N$ theory.
A gauge theory description is given by the quiver~\cite{Bergman:2014kza,Hayashi:2014hfa}
\begin{align}
\label{TNquiver}
[2]\stackrel{y_1}{-}(2)\stackrel{x_1}{-}(3)-\cdots-(N-2)\stackrel{x_{N-3}}{-}(N-1)\stackrel{y_2}{-}[N] \,.
\end{align}
That is, a linear quiver with $\SU(K)$ gauge groups whose rank increases from $1$ to $N-2$ and which has two hypermultiplets in the fundamental representation of $\SU(2)$ and $N$ hypermultiplets in the fundamental representation of $\SU(N-1)$. Between adjacent gauge group nodes there are once again bi-fundamental hypermultiplets denoted by $x_i$. For $N=3$ this is the rank-$1$ Seiberg theory with global symmetry enhanced to $E_6$.

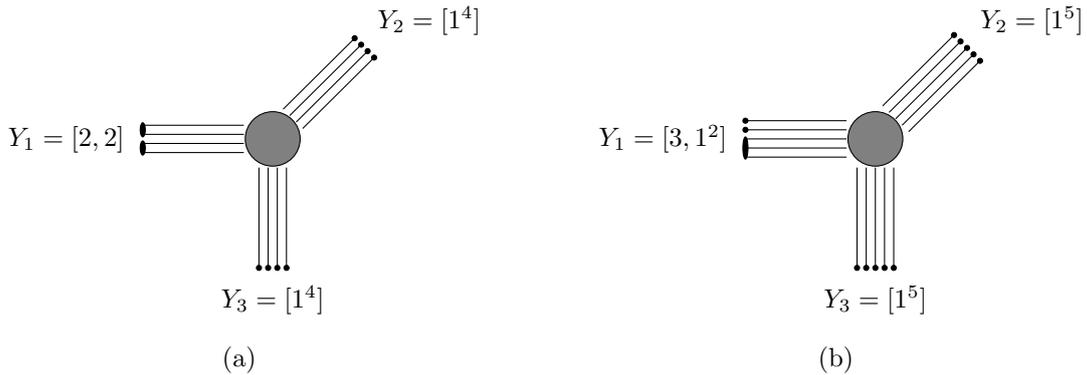
\begin{figure}
\centering
\subfigure[][]{\label{fig:TN-web-2}
\begin{tikzpicture}[scale=0.95]
    \foreach \i in {-3/2,-1/2,1/2,3/2}{
      \draw (-0.4,0.127*\i) -- (-1.8,0.127*\i);
      \draw (0.127*\i,-0.4) -- (0.127*\i,-1.8) [fill=black] circle (1pt);
      \draw (0.28+0.09*\i,0.28-0.09*\i) -- (1.27+0.09*\i,1.27-0.09*\i) [fill=black] circle (1pt) ;
    }
    \draw[fill=black] (-1.8,0.127) ellipse (1pt and 2.8pt);
    \draw[fill=black] (-1.8,-0.127) ellipse (1pt and 2.8pt);
    \draw[fill=gray] (0,0) circle (0.38);
    
    \node [anchor=east] at (-1.9,0) {\footnotesize $Y_1=[2,2]$};
    \node [anchor=south west] at (1.3,1.3) {\footnotesize $Y_2=[1^4]$};
    \node [anchor=north] at (0,-1.9) {\footnotesize $Y_3=[1^4]$};
\end{tikzpicture}
}\hskip 10mm
\subfigure[][]{\label{fig:TN-web-3}
\begin{tikzpicture}[scale=0.95]
    \foreach \i in {-2,-1,0,1,2}{
      \draw (-0.4,0.127*\i) -- (-1.8,0.127*\i);
      \draw (0.127*\i,-0.4) -- (0.127*\i,-1.8) [fill=black] circle (1pt);
      \draw (0.28+0.09*\i,0.28-0.09*\i) -- (1.27+0.09*\i,1.27-0.09*\i) [fill=black] circle (1pt) ;
    }
    \draw (-1.8,0.127*2) [fill=black] circle (1pt);
    \draw (-1.8,0.127*1) [fill=black] circle (1pt);
    \draw[fill=gray] (0,0) circle (0.38);
    \draw[fill=black] (-1.8,-0.127*1) ellipse (1pt and 4.4pt);

    \node [anchor=east] at (-1.9,0) {\footnotesize $Y_1=[3,1^2]$};
    \node [anchor=south west] at (1.3,1.3) {\footnotesize $Y_2=[1^5]$};
    \node [anchor=north] at (0,-1.9) {\footnotesize $Y_3=[1^5]$};
\end{tikzpicture}
}
\caption{$T_{N,K,j}$ junctions: (a) $N=4$, $K=j=2$ and (b) $N=5$, $j=1$, $K=3$. \label{fig:TN-webs}}
\end{figure}

\paragraph*{$T_{N,K,j}$ theory:} Theories with multiple 5-branes ending on the same 7-brane can be obtained from the unconstrained $T_N$ theories by Higgs branch flows, and in general have reduced global symmetries. 
For the theories we will consider, the NS5 and $(1,1)$ 5-branes are each ending on an individual 7-brane. The D5-branes are split into $j$ groups of $K>1$ D5-branes, each terminating on a single D7-brane, and $N-jK$ unconstrained D5-branes (see figure~\ref{fig:TN-webs}). This corresponds to the partitions
\begin{align}
 Y_1& \ = \ [K^j,1^{N-K j}] \, , \qquad Y_2 \ = \ Y_3 \ = \ [1^N] \, .
\end{align}
The global symmetry for generic $N$, $K$, $j$ is reduced from 
$\SU(N)^3$ to $\SU(N-j K)\times \SU(j)\times \SU(N)^2\times \U(1)$. The $T_{4,2,2}$ junction realizes the rank-$1$ Seiberg theory with global symmetry enhanced to $E_7$.

Gauge theory deformations can be read off conveniently after deforming the brane web as discussed in~\cite{Chaney:2018gjc}.
For $N>jK$ this yields
\begin{align}\label{eq:TNKj-quiver-1}
[N-jK]\stackrel{y_1}{-}(N-jK+j-1)\stackrel{x_1}{-}
\dots\stackrel{x_{K-1}}{-}(N&-K)\stackrel{x_K}{-}
\cdots \stackrel{x_{N-3}}{-}(2)\stackrel{y_3}{-}[2] \,.\nonumber \\
&\ \,| \, {\scriptstyle y_2}\\
&\,[j] \nonumber
\end{align}
Between the links labeled by $x_{K}$ and $x_{N-3}$ the rank of the gauge groups decreases in steps of one.
There are $K-2$ gauge nodes between the links labeled by $x_1$ and $x_{K-1}$, with rank increasing in steps of $j-1$. 
For $j=1$ there is a total of $K$ $\SU(N-K)$ gauge nodes. 

For $N=jK$, when there are no unconstrained D5-branes, the form of the gauge theory deformation depends on whether $j=2$ or $j>2$. For $j=2$ and $N=2K$ the gauge theory is given by
\begin{align}\label{eq:TNKj-quiver-2}
 [2]\stackrel{y_1}{-}(2)\stackrel{x_1}{-}(3)-\cdots - (K-1)\stackrel{x_{K-1}}{-} (&K) \stackrel{x_K}{-} (K-1) -\cdots -(3)\stackrel{x_{N-3}}{-}(2)\stackrel{y_3}{-}[2] \,.\nonumber\\
 &\,|\,{\scriptstyle y_2} \\
 &\,\!\![2]\nonumber
\end{align}
The quiver is symmetric under reflection across the $\SU(K)$ node and may be seen as a gluing of two $T_K$ theories, gauging their respective $\SU(K)$ flavor symmetries and adding two fundamental hypermultiplets at the central node.

For $j>2$ and $N=jK$ the gauge theory deformation is given by the quiver
\begin{align}\label{eq:TNKj-quiver-3}
 (j-1)\stackrel{x_1}{-}(2(j-1))-\cdots  \stackrel{x_{K-1}}{-} (N&-K) \stackrel{x_{K}}{-} (N-K-1) -\cdots \stackrel{x_{N-3}}{-}(2)\stackrel{y_3}{-}[2] \,.\nonumber\\
 &\, \ | \,{\scriptstyle y_2}\\
 &\, \ \!\![2]\nonumber
\end{align}
There are $K-2$ gauge groups between the bi-fundamental fields $x_1$ and $x_{K-1}$, with rank increasing in steps of $j-1$,
and $(N-K-3)$ gauge groups between the bi-fundamental fields $x_K$ and $x_{N-3}$ with rank decreasing by one.

\subsection{The topologically twisted index}

The topologically twisted index~\cite{Benini:2015noa} of a five-dimensional $\cN = 1$ theory is the (Euclidean) partition function on $\cM_4 \times S^1$, with a partial topological twist on $\cM_4$, and can be computed using localization~\cite{Hosseini:2018uzp}. Such computations were performed for $\cM_4$ either a toric K\"ahler manifold~\cite{Hosseini:2018uzp} or a product of two Riemann surfaces~\cite{Hosseini:2018uzp,Crichigno:2018adf}, \ie\;$\cM_4=\Sigma_{\fg_1} \times \Sigma_{\fg_2}$, where $\fg_1$ and $\fg_2$ denote the genus of the respective complex curve.\footnote{As a property of the partial topological twist, geometrically, only the genus of the respective curve is relevant for the computation of the twisted index.}

Upon localization, the twisted index evaluates to a function of flavor magnetic fluxes $\fs_I$, parameterizing the twist, and fugacities $y_I = e^{i \beta \Delta_I}$ for the flavor symmetries of the theory.
Alternatively, the partition function can be viewed as the trace over the Hilbert space $\cH_{\cM_4}$ of states in radial quantization on $\cM_{4}$, \emph{i.e.}
\be
 Z ( y_I , \fs_I )  \ = \  \Tr_{\cH_{\cM_4}} (-1)^F e^{- \beta H} \prod_{I} y_I^{J_I} \,.
\ee
This is the equivariant Witten index of the dimensionally reduced quantum mechanics~\cite{Witten:1982df}, where the fluxes explicitly enter in the Hamiltonian $H$, and $J_I$ are the generators of the flavor symmetries.

\subsubsection{Matrix model}

As derived in~\cite{Hosseini:2018uzp}, the localized partition function can be written as a contour integral%
\footnote{The contour $\cC$ ought to be dictated by supersymmetric localization; however, an \emph{ab initio} derivation of its form has not been performed as of now (see \emph{e.g.}~\cite{Hosseini:2018uzp}).
For the purpose of this paper, the detailed contour will be irrelevant, because we rely on an alternative description of the twisted index, as given in section~\ref{bethesum}.}
\be
 \label{5D:twisted:index:general}
 Z_{\cM_4 \times S^1} (\fs , y , q)  \ = \  \frac{1}{|\mathfrak{W}|} \sum_{k = 0}^{\infty} \sum_{\{\fp\} \in \Gamma_\fh} \oint_\cC \left[ \prod_{x \in \mathbb{T}} \frac{\diff x}{2\pi \ii x} \right]q^k Z_{\text{int}}^{(k\text{-instantons})} (\fp, x; \fs_I, y_I, q) \,,
\ee
of a meromorphic function in variables $x_\ell = e^{\ii \beta a_\ell}$, where $a_\ell$ are the Coulomb branch parameters of the four-dimensional theory obtained by reducing the five-dimensional theory along $S^1$.
The sum over $\fp$ is over a set of gauge magnetic fluxes $\fp$, living in the co-root lattice $\Gamma_{\fh}$ of the gauge group $G$.%
\footnote{We are interested in the ``non-equivariant" limit of the topologically twisted index, \ie~$\epsilon_{i} \to 0$, $i=1,2$,
where $\epsilon_i$ are the $\Omega$-deformation parameters.
Thus, we shall consider the sum in \eqref{5D:twisted:index:general} over \emph{topological} fluxes. We refer the reader to~\cite{Hosseini:2018uzp} for more details.}
Here $q = e^{- 8 \pi^2 \beta / g_{\text{YM}}}$, where $g_{\text{YM}}$ is the five-dimensional gauge coupling constant, and $\beta$ is the circumference  of $S^{1}$. We also denoted the order of the Weyl group of $G$ by $|\mathfrak{W}|$. Finally, the function $Z_{\text{int}}^{(k\text{-instantons})} (\fp, x; \fs_I, y_I, q)$ receives contributions from the classical action, the one-loop determinants, and the instantons. 

For the purpose of this paper, we work in a strict large $N$ limit, in which we expect instantonic contributions (\emph{i.e.}\;$k>0$ in~\eqref{5D:twisted:index:general}) to be suppressed~\cite{Jafferis:2012iv}, and thus we shall solely work with the ``perturbative" part, and write $Z_{\text{int}}  \ \equiv \ Z_{\text{int}}^{(0\text{-instantons})} (\fp, x; \fs_I, y_I, q) $. Then, for a theory of gauge group $G$, coupled to a set of hypermultiplets in a representation $\oplus_{I} (\fR_I \oplus \bar \fR_I)$, and for $\cM_4 = \Sigma_{\fg_1} \times \Sigma_{\fg_2}$, it reads~\cite{Hosseini:2018uzp,Crichigno:2018adf}
\bea
 \label{index:Sigmag1xSigmag2xS1}
 Z_{\text{int}}  \ = \ &  \bigg( \det\limits_{\ell m} \frac{\partial^2 \wt \cW (a , \fn)}{\partial a_\ell \partial a_m} \bigg)^{\fg_1}
 e^{\frac{8 \pi^2 \beta}{g_{\text{YM}}^2} \Tr_{\text{F}} ( \fm \fn ) + k \beta \Tr_{\text{F}} (\fm \fn a) }
 \prod_{\alpha \in G} \bigg( \frac{1 - x^\alpha}{x^{\alpha/2}} \bigg)^{(\alpha(\fm)+ 1 - \fg_1 ) ( \alpha(\fn)+ 1 - \fg_2 )} \\ &
 \times \prod_{I} \prod_{\rho_I \in \fR_I} \bigg( \frac{x^{\rho_I/2} y^{\nu_I/2}}{1 - x^{\rho_I} y^{\nu_I}} \bigg)^{( \rho_I(\fm) + \nu_I(\fs) + \fg_1 - 1 ) ( \rho_I(\fn)  + \nu_I(\ft) + \fg_2 - 1 )} \, ,
\eea
where $\alpha$ denote the roots of $G$, and $\rho_I$, $\nu_I$ are the weights of the hypermultiplets under the gauge, flavor symmetry group, respectively.
Moreover, $k$ is the Chern-Simons level, $(\fm , \fn)$, $(\fs_I , \ft_I)$ are the gauge, background magnetic fluxes on $(\Sigma_{\fg_1},\Sigma_{\fg_2})$, respectively. Furthermore, by $\Tr_{\text{F}}$, we denote the trace in the fundamental representation. An important ingredient entering~\eqref{index:Sigmag1xSigmag2xS1} is the effective twisted superpotential of the two-dimensional topological field theory on $\Sigma_{\fg_1}$, which receives contributions from the Kaluza-Klein modes on $\Sigma_{\fg_2} \times S^1$. It is explicitly given by~\cite{Hosseini:2018uzp,Nekrasov:2014xaa}
\be
 \label{pert:twisted:superpotential}
 \begin{aligned}
  \wt \cW (a , \fn ; & \Delta , \ft)  \ = \  - \frac{8 \pi^2 \ii \beta}{g_{\text{YM}}^2} \Tr_{\text{F}} (\fn  a) + \frac{k \beta}{2} \Tr_{\text{F}} (\fn a^2) \\ &
  + \frac{1}{\beta} \sum_{\alpha \in G} \left( \alpha (\fn) + 1 - \fg_2 \right) \left[ \Li_2 ( x^\alpha ) - \frac{1}{2} g_2 \left( - \alpha(\beta a) \right) \right] \\ &
  - \frac{1}{\beta} \sum_I \sum_{\rho_I \in \fR_I} \left( \rho_I(\fn) + \nu_I (\ft) + \fg_2 - 1 \right) \left[ \Li_2 ( x^{\rho_I} y^{\nu_I} ) - \frac{1}{2} g_2 \left( \rho_I(\beta a) + \nu_I (\beta \Delta) \right) \right] ,
 \end{aligned}
\ee
where $\Li_s (x)$ are the polylogarithm functions, and
\bea
 \label{g:functions}
 g_1(a)     \  =  \ & \  a - \pi \,,\\
 g_2 ( a )  \  =  \ & \  \frac{a^2}{2} - \pi a + \frac{\pi^2}{3} \,, \\
 g_3 ( a )  \  =  \ & \  \frac{a^3}{6} - \frac{\pi}{2} a^2 + \frac{\pi^2}{3} a \,.
\eea
The functions $ g_s ( a )$ satisfy the following identity
\be \label{Z2:gs}
 g_s (2 \pi - a)  \ = \  (-1)^s g_s(a) \, .
\ee
Assuming $0 < \Delta < 2 \pi$, and $t \in \mathbb{R}$, the polylogarithm asymptotically simplifies as follows
\be \label{PolyLog:asymptotic}
 \Li_s (e^{t + \ii \Delta}) \ \sim \  \ii^{s-2} g_s( - \ii t + \Delta ) \, , \qquad  t \to \infty \,.
\ee

\subsubsection{Bethe sum formula}\label{bethesum}

One of the technical challenges of our work is the explicit evaluation of the five-dimensional topologically twisted index on $\Sigma_{\fg_2} \times ( \Sigma_{\fg_1} \times S^1 )$ in the large $N$ limit.
We shall use the strategy employed in~\cite{Hosseini:2018uzp}. Namely, we first exchange the sum over the magnetic flux lattice, $\fm \in \Gamma_\fh$, with the contour integral in~\eqref{5D:twisted:index:general}. Consequently, we can rewrite the twisted index as follows%
\footnote{Here, the sum $\sum_{\fm}$ is over a wedge $\Gamma_{\fh}^\cC$ inside the magnetic lattice, for which $Z_{\text{int}} (a,\fm,\fn)$ has poles inside the contour.}
\be
 \label{5d:3d:index:intro}
 \sum_{\fn \in \Gamma_\fh} \oint_{\cC} \sum_{\fm \in \Gamma_\fh^{\cC}} Z_{\text{int}} (a,\fm,\fn) \ = \
 \sum_{\fn \in \Gamma_\fh} \oint_{\cC} \sum_{\fm \in \Gamma_\fh^{\cC}} e^{\ii \fm_\ell \frac{\partial \wt \cW(a, \fn)}{\partial a_\ell}} Z_{\text{int}}(a,\fm=0,\fn) \, .
\ee
Thus, we may interpret the five-dimensional partition function as arising from a three-dimensional theory on $\Sigma_{\fg_1} \times S^1$ summed over topological sectors on $\Sigma_{\fg_2}$, which are labeled by $\fn \in \Gamma_\fh$.
The sum over gauge fluxes $\fm$ in~\eqref{5d:3d:index:intro} is a geometric series and so can be done straightforwardly.%
\footnote{We take a large positive integer cut-off $P$, and then perform the summation over $\fm \leq P-1$ in~\eqref{5d:3d:index:intro}. The final result is independent of $P$ because of~\eqref{BAEs:general}.}
We shall then, using the results of~\cite{Benini:2015eyy}, set up an auxiliary large $N$ problem for finding the positions of the poles of the contour integral \eqref{5d:3d:index:intro}
by solving the so-called Bethe ansatz equations (BAEs)~\cite{Hosseini:2018uzp,Nekrasov:2014xaa}
\be \label{BAEs:general}
 \exp \bigg( \ii \frac{\partial \wt\cW (a , \fn ; \Delta , \ft)}{\partial a_\ell} \bigg) \bigg|_{a = a_{(\ell)}} \ = \  1 \, , \qquad \ell  \ = \  1, \ldots , \text{rk}(G) \, .
\ee
Therefore, these BAEs~\eqref{BAEs:general} determine the Coulomb branch parameters $a_{(\ell)}$ in terms of the gauge fluxes $\fn_\ell$.
Hence, we need an additional set of equations in order to fix both $a_{(\ell)}$ as well as $\fn_\ell$ uniquely.
Following~\cite{Hosseini:2018uzp}, we \emph{propose} that the correct condition to be imposed at large $N$, is given by%
\footnote{The relevance of the effective Seiberg-Witten prepotential in this context has been suggested also in \cite{Crichigno:2018adf}. However, the gauge fluxes were set to zero from the outset in the proposal of \cite{Crichigno:2018adf}. The results were found to disagree with the supergravity computations in Type IIA, and we find similar mismatches with Type~IIB supergravity for the dualities considered here.}
\be
 \label{MAEs:general}
\exp \bigg( \frac{2 \pi \ii}{\hbar} \frac{\partial \cF(a)}{\partial a_\ell} \bigg) \bigg|_{a = a_{(\ell)}} \ = \  1 \, , \qquad \ell  \ = \  1, \ldots , \text{rk}(G) \, ,
\ee
where $\hbar = \epsilon_1$ is the $\Omega$-background parameter and $\cF (a)$ is the effective Seiberg-Witten prepotential of the four-dimensional theory, obtained by compactifying the five-dimensional theory on a circle whilst retaining the Kaluza-Klein modes~\cite{Nekrasov:1996cz}. It reads
\be \label{pert:prepotential}
\begin{aligned}
  2 \pi \ii \cF (a ; \Delta) \ = \ &   - \frac{4 \pi^2 \beta}{g_{\text{YM}}^2} \Tr_{\text{F}} (a^2) - \frac{\ii k \beta}{6} \Tr_{\text{F}} (a^3) \\ &
  - \frac{1}{\beta^2} \sum_{\alpha \in G} \left[ \Li_3 ( x^\alpha ) + \frac{\ii}{2} g_3 \left( - \alpha( \beta a ) \right) - \zeta(3) \right] \\ &
  + \frac{1}{\beta^2} \sum_{I} \sum_{\rho_I \in \fR_I} \left[ \Li_3 ( x^{\rho_I} y^{\nu_I} ) - \frac{\ii}{2} g_3 \left( \rho_I( \beta a ) + \nu_I ( \beta \Delta ) \right) - \zeta(3) \right] \,,
 \end{aligned}
\ee
where $g_3(a)$ is given in~\eqref{g:functions}, and the remaining notation is understood as before.

The solutions to equations~\eqref{BAEs:general} and~\eqref{MAEs:general} in the large $N$ limit are then used to evaluate the topologically twisted index using the residue theorem.
We thus obtain an alternative description of the matrix model~\eqref{index:Sigmag1xSigmag2xS1} as follows\footnote{Notice that the exponent of the determinant factor is changed due to the addition of the Jabobian for the change of variables.}
\bea \label{Bethe:sum:Z}
 & Z_{ \Sigma_{\fg_2} \times ( \Sigma_{\fg_1} \times S^1)} (\fs , \ft , \Delta)  \ = \ 
 \frac{(-1)^{\text{rk}(G)}}{|\mathfrak{W}|} \sum_{\fn \in \Gamma_\fh} \sum_{a = a_{(\ell)}} Z_{\text{int}} \big|_{\fm = 0} (a , \fn)
 \bigg( \det\limits_{\ell m} \frac{\partial^2 \wt \cW (a , \fn)}{\partial a_\ell \partial a_m} \bigg)^{\fg_1- 1} \,.
\eea

In the following sections we perform the universal topological twist \cite{Benini:2015bwz} by setting
\be
 \Delta \ = \ \pi \, , \qquad \fs \ = \ 1 - \fg_1 \, , \qquad \ft \ = \ 1 -\fg_2 \, .
\ee
This can be done \emph{only} for Riemann surfaces of negative curvature, \ie\;$\fg_1>1$ and $\fg_2>1$.

We discuss the explicit form of the matrix models computing the topologically twisted index for the $T_N$, $\#_{N,M}$, and $T_{N,K,j}$ theories in appendix~\ref{app:matrix-models}.

\subsection{Numerical methodology}
\label{sec:numerics}

We now briefly review our methodology for computing the numerical large $N$, $M$ partition functions.
To do so, we first recall two standard arguments, and then proceed to outlining our numerical saddle point method, which we use to get our explicit numerical data.

Firstly, let us remark that for both the topologically twisted index and the five-sphere partition function, we assume that the ultraviolet superconformal fixed point partition functions are captured by their infrared gauge theory description, which we reviewed in section~\ref{sec:quivers} for the theories at hand. This is reliant on the conjecture that higher derivative terms required to describe the infrared superconformal fixed points are $\cQ$-exact, and thus do not contribute to the partition function computed with respect to the same supercharge $\cQ$~\cite{Jafferis:2012iv,Kim:2012ava,Kim:2012qf}.

Secondly, we shall in the following assume that the instantonic contributions to both the topologically twisted index and five-sphere partition function are suppressed in the large $N$, $M$ limit. It was argued in~\cite{Jafferis:2012iv}, that (as we move onto the Coulomb branch) the contributions arising from the localization in an instanton background are effectively dependent on the exponential of the Coulomb branch parameters $\sim e^{-|a |}$, where we collectively denote the Coulomb branch parameters by $a$. For the case at hand, it can explicitly be checked, that the $a$'s scale as $N^{\alpha}$, $N^{\alpha} M^{\beta}$, with $\alpha >0$, for $T_{N,k,j}$ (including $T_N$), $\#_{N,M}$ theories, respectively. Consequently, we expect that only the zero-instanton -- or perturbative -- part of the partition functions contribute in the large $N$, $M$ limit, which simplifies our calculations considerably.

To evaluate the large $N$, $M$ perturbative partition functions, we employ the numerical saddle point approximation~\cite{Herzog:2010hf}: for the five-sphere partition functions as well as the topologically twisted indices, we are ultimately required to extremize some function $f(a_i)$ with respect to a set of parameters $\left\{ a_i \right\}_{i \in I}$, where $I$ is an index set which varies on a case-by-case basis. Hence, we would like to solve
\be\label{extr}
\frac{\partial f}{\partial a_{i}} \ = \ 0 \,, \qquad \forall i \in I \,.
\ee
In general, these equations are non-trivial. To solve them, we reinterpret these equations as describing a system of $|I|$ particles with time-dependent coordinates $a_i(t)$, moving in a potential given by the function $f$. Then, the resulting equations of motion read
\be
\frac{\partial f}{\partial a_{i}(t)} \ = \ \xi \frac{\diff a_i(t)}{\diff t} \,, \qquad \forall i \in I\,,
\ee
where $\xi \in \{\pm 1\}$ has to be fixed such that the potential is attractive. The explicit choice of $\xi$ depends on the theory and the partition function in question. Then, at large time $t \to \infty$, we approach an equilibrium configuration, which describes the solutions to the extremization equations~\eqref{extr}.

Notice, that in the case of the five-sphere partition function, we arrive at the equations~\eqref{extr} as the saddle point equations for the integrand of $-\log Z^{S^5}_{\rm pert}$, \ie\;the free energy of the theory. By solving the extremization equations, we solve for the saddle points which govern (at leading order in $N$, $M$) the Coulomb branch integral as given in~\eqref{S5pert}.\footnote{We also refer to~\cite{Fluder:2018chf} for more details on this case.} 

In the case of the topologically twisted index, we employ the (conjectured) ``Bethe sum formula", as explained in section~\ref{bethesum}.
Thus, we first solve the equations~\eqref{MAEs:general} for the Coulomb branch parameters $a_i$, using the numerical saddle point method. In a second step, given the solutions for $a_i$, we solve the BAEs for the (perturbative) twisted superpotential $\widetilde \cW$, given in equation~\eqref{BAEs:general}, for the gauge magnetic fluxes $\fn_{i}$, \ie~now the fluxes $\fn_{i}(t)$ describe the worldlines of the particles in the potential given by $f \equiv \widetilde\cW$ in~\eqref{extr}. The $\fn_i$'s being large in the large $N$, $M$ limit, we can effectively consider them as continuous variables, and thus their solutions will in general neither be integer nor real.

Finally, given the numerical solutions of the corresponding BAEs for the Coulomb branch parameters $a_i$, and the gauge fluxes $\fn_i$,
we can evaluate the topologically twisted index using the Bethe sum formula~\eqref{Bethe:sum:Z}, where  one can explicitly check that the contribution of the determinant in~\eqref{Bethe:sum:Z}, \ie\;
\be
 \bigg( \det\limits_{\ell m} \frac{\partial^2 \wt \cW (a , \fn)}{\partial a_\ell \partial a_m} \bigg)^{\fg_1- 1}\,,
\ee
is suppressed in the large $N$, $M$ limit.

We remark that we use the asymptotic versions of the effective Seiberg-Witten prepotential, twisted superpotential and twisted index as detailed in appendix~\ref{app:matrix-models}.
This is sufficient to determine the first few leading orders in the large $N$, $M$ limit of the twisted index, and substantially speeds up the numerical evaluation.

Lastly, let us comment on the region of validity of this method.
Of course, the saddle point approximation can only be assumed to be rigorous in the strict large $N$, $M$ limit.
At subleading orders, we expect that other saddle points will contribute to the integral, and, eventually, instanton contributions will be important.
For instance, in the case of the twisted index, at finite $N$, one is required to sum over many Bethe vacua, as opposed to only considering the dominant solution to~\eqref{BAEs:general} and~\eqref{MAEs:general}.

\begin{table}
\centering
\setlength{\tabcolsep}{7.5pt}
\renewcommand{\arraystretch}{1.2}
\begin{tabular}{c|c||c|c||c|c}
\toprule
 $N$ & $F_{S^5}(T_{N,N/2,1})/N^4$ & $N$  & $ F_{S^5}(T_{N,N/2,2})/N^4$ & $N$ & $F_{S^5}(T_{N,N/4,3})/N^4$ \\
\hline
$20$ & $0.231123$ & $ 20 $ & $ 0.128039 $   & $8$  & $0.282851$ \\
$22$ & $0.232605$ & $ 22 $ & $ 0.129099 $   & $12$ & $0.295506$\\
$24$ & $0.233815$ & $ 24 $ & $ 0.129974 $   & $16$ & $0.301709$\\
$26$ & $0.234819$ & $ 26 $ & $ 0.130710 $   & $20$ & $0.305356$\\
$28$ & $0.235665$ & $ 28 $ & $ 0.131338 $   & $24$ & $0.307747$\\
$30$ & $0.236387$ & $ 30 $ & $ 0.131879 $   & $28$ & $0.309431$\\
$32$ & $0.237011$ &  $32$  & $0.132351$     & $32$ & $0.310679$\\
$34$ & $0.237555$ &  $34$  & $0.132766$     & $36$ & $0.311640$\\
$36$ & $0.238032$ &  $36$  & $0.133133$     & $40$ & $0.312401$\\
$38$ & $0.238455$ &  $38$   & $0.133461$    & $44$ & $0.313020$\\
$40$ & $0.238832$ &  $40$   & $0.133756$    & $48$ & $0.313531$\\
\bottomrule
\end{tabular}
\caption{Numerical values for the saddle point evaluation of the five-sphere free energy for $T_{N,K,j}$ theories.}\label{tbl:TNKj-1}
\end{table}

\subsection{Large \texorpdfstring{$N$}{N} results}

We start with the free energy of the five-sphere partition function
\be
F_{S^5}(T_{N,K,j}) \ = \ - \log  Z_{\rm pert}^{S^5}(T_{N,K,j})
\ee
 for a sample of $T_{N,K,j}$ theories. For $N=2K$ with $j=2$, $N=2K$ with $j=1$, and $N=5K$ with $j=3$, the data are shown in table~\ref{tbl:TNKj-1}.
 The leading large $N$ behavior extracted from this data by fitting to a polynomial is
\bea
 F_{S^5} (T_{N,N/2,1})& \ = \ 0.24530\,N^4\,,\\
 F_{S^5} (T_{N,N/2,2})& \ = \ 0.13924\,N^4\,,\\
 F_{S^5} (T_{N,N/5,3})& \ = \ 0.31900\, N^4\,.
\eea
This agrees at the per mille level with the supergravity predictions \eqref{eq:FS5-sugra-TNKj}. We note that the results are highly sensitive towards the quiver data; for instance, the fundamental hypermultiplets denoted by $y_i$ in the quiver gauge theories \eqref{eq:TNKj-quiver-1}, \eqref{eq:TNKj-quiver-2}, \eqref{eq:TNKj-quiver-3} affect the leading large $N$ result non-trivially, despite being subleading compared to the bi-fundamental hypermultiplets and gauge fields by free field counting arguments.

\begin{figure}
 \centering
 \subfigure[][]{\label{fig:largeN-TN}
 \includegraphics[width=0.45\linewidth]{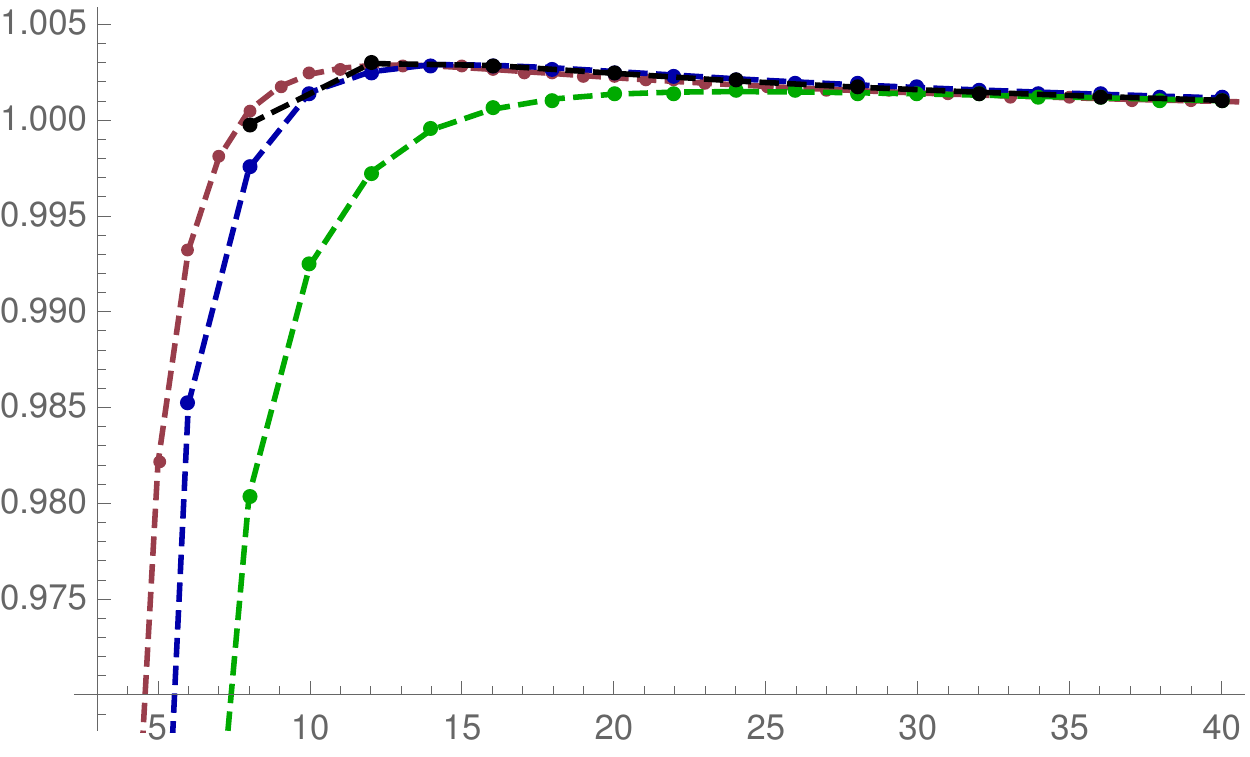}
 \put (0,-1) {\footnotesize $N$}
 \put (-210,140) {\footnotesize $-\frac{9}{8}\frac{\log Z}{(1-\fg_1)(1-\fg_2)F_{S^5}}$}
 }\qquad
 \subfigure[][]{\label{fig:largeN-HT}
 \includegraphics[width=0.45\linewidth]{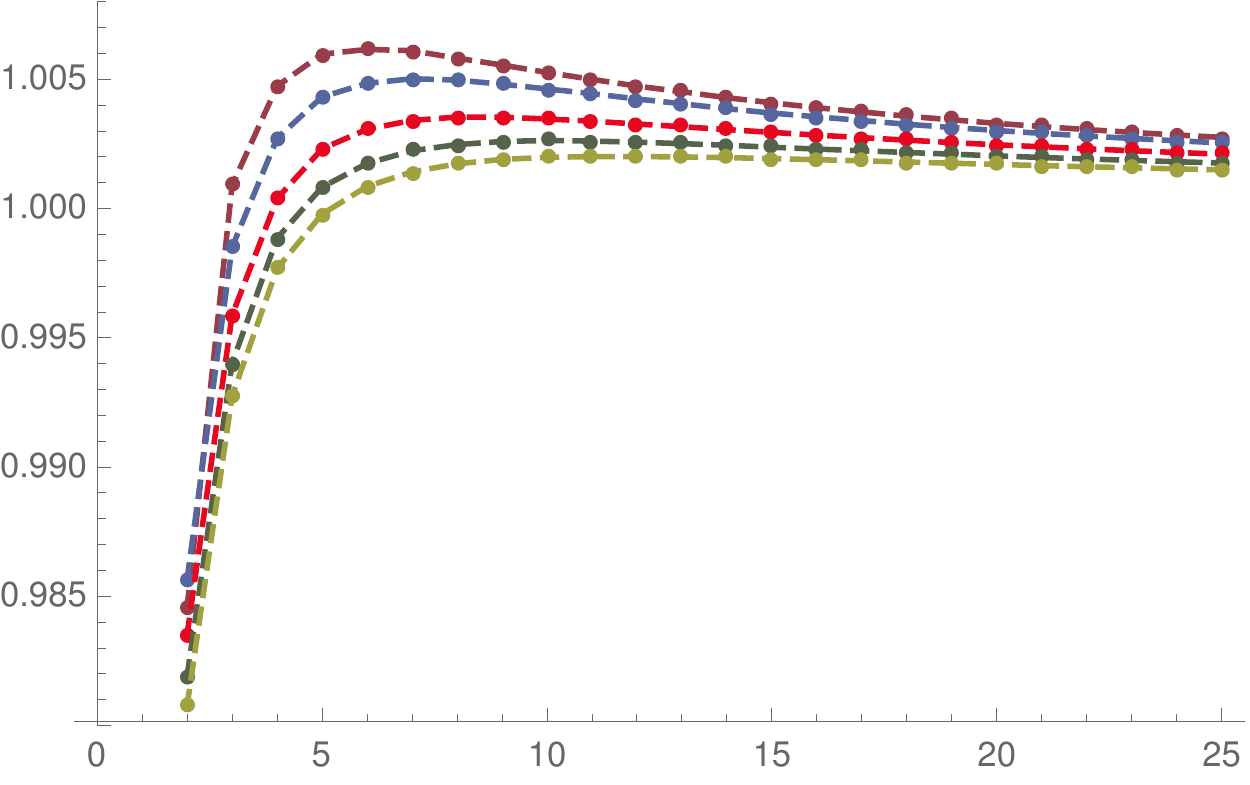}
 \put (0,-1) {\footnotesize $M$}
 \put (-210,140) {\footnotesize $-\frac{9}{8}\frac{\log Z}{(1-\fg_1)(1-\fg_2)F_{S^5}}$}
 }
 \caption{Ratio of the topologically twisted index and the five-sphere partition function, evaluated numerically as described in section~\ref{sec:numerics}.
 On the left hand side for the $T_N$ (red), $T_{N,N/2,1}$ (blue), $T_{N,N/2,2}$ (green) and $T_{N,N/4,3}$ (black) theories.
 On the right hand side for $\#_{N,M}$ as functions of $M$. From top to bottom the curves are $N\in\lbrace 5,10,15,20,25\rbrace$. As function of $N$ the ratio exhibits similar convergence properties.
 \label{fig:plot-1}}
\end{figure}

The results of the numerical evaluation of the ratio of the topologically twisted index and the five-sphere partition function for the $T_N$ theories and the sample of $T_{N,K,j}$ theories are shown in figure~\ref{fig:largeN-TN}, and for the $\#_{N,M}$ theories in figure~\ref{fig:largeN-HT}.
The sphere partition functions for the $T_{N,K,j}$ theories are taken from table~\ref{tbl:TNKj-1}, and for the $T_N$ and $\#_{N,M}$ theories  from \cite{Fluder:2018chf}.
The results clearly show that the ratio approaches the universal value for large $N$ for the $T_N$ and $T_{N,K,j}$ theories,  and for large $N$ and $M$ for the $\#_{N,M}$ theories. In fact, the ratio is well captured by the large $N$ asymptotics significantly earlier than the individual quantities. This suggests that the leading $1/N$ corrections to the individual quantities, that remain after using the large $N$ approximations we have employed, cancel in the ratio.
Moreover, for the $\#_{N,M}$ theories the ratio appears to approach the universal value if either $N$ or $M$ are large. 
This is in contrast to the sphere partition function, which approaches the supergravity result only if $N$ \emph{and} $M$ are large~\cite{Fluder:2018chf}.

\section{Magnetic \texorpdfstring{AdS$_6$}{AdS6} black holes in Type IIB}\label{sec:sugra}

In this section we discuss AdS$_6$ black holes with horizon topology $\Sigma_{\fg_1}\times\Sigma_{\fg_2}$ in Type IIB and their Bekenstein-Hawking entropies.
The starting point are the ${\rm AdS}_6\times S^2\times\Sigma$ solutions to type IIB supergravity of~\cite{DHoker:2016ujz,DHoker:2016ysh,DHoker:2017mds,DHoker:2017zwj}, characterized by a choice of Riemann surface $\Sigma$ and two locally holomorphic functions $\cA_\pm$. The consistent Kaluza-Klein reduction to six-dimensional $F(4)$ gauged supergravity for this general class of solutions~\cite{Hong:2018amk,Malek:2018zcz}
allows to uplift the six-dimensional AdS$_2 \times\Sigma_{\fg_1}\times\Sigma_{\fg_2}$ solutions of~\cite{Suh:2018tul} to Type IIB.
For each AdS$_6$ solution in Type IIB, the uplift produces a distinct black hole solution. In the following we introduce the relevant background. We also compute the sphere partition functions for the $T_{N,K,j}$ theories.

\subsection{\texorpdfstring{${\rm AdS}_6$}{AdS6} solutions in Type IIB}\label{sec:ads6}

With a complex coordinate $w$ on $\Sigma$, the metric and complex two-form in the ${\rm AdS}_6\times S^2\times\Sigma$ solutions of~\cite{DHoker:2016ujz,DHoker:2016ysh,DHoker:2017mds,DHoker:2017zwj} are given by
\begin{align}\label{eqn:ansatz}
 \diff s^2  \ = \  f_6^2 \, \diff s^2 _{\mathrm{AdS}_6} + f_2 ^2 \, \diff s^2 _{S^2} + 4\rho^2 \diff w \diff \bar w \, ,
 \qquad
 C_{(2)} \ = \ \cC \vol_{S^2} \,,
\end{align}
where $\vol_{S^2}$ is the canonical volume form on a two-sphere, $S^2$, with unit radius. The solutions are parametrized by two locally holomorphic functions $\cA_\pm$ on $\Sigma$, in terms of which the metric functions are
\begin{align}\label{eqn:metric}
f_6^2& \ = \ \sqrt{6\cG T} \, ,
&
f_2^2& \ = \ \frac{1}{9}\sqrt{6\cG}\,T ^{-\tfrac{3}{2}} \, ,
&
\rho^2& \ = \ \frac{\kappa^2}{\sqrt{6\cG}} T^{\tfrac{1}{2}} \, ,
\end{align}
where
\begin{align}\label{eq:kappa-G}
 \kappa^2 & \ = \ -|\partial_w \cA_+|^2+|\partial_w \cA_-|^2 \, ,
 &
 \partial_w\cB & \ = \ \cA_+\partial_w \cA_- - \cA_-\partial_w\cA_+ \, ,
 \\
 \cG& \ = \ |\cA_+|^2-|\cA_-|^2+\cB+\bar{\cB} \, ,
 &
  T^2& \ = \ \left(\frac{1+R}{1-R}\right)^2 \ = \ 1+\frac{2|\partial_w\cG|^2}{3\kappa^2 \, \cG } \, .
  \label{eqn:Gdef}
\end{align}
The function $\cC$ and the axion-dilaton scalar $B=(1+\ii\tau)/(1-\ii\tau)$ are given by
\begin{align}\label{eqn:flux}
 \mathcal{C} &  \ = \   \frac{2\ii}{3}\left(
 \frac{\partial_{\bar w}\cG\partial_w\cA_++\partial_w \cG \partial_{\bar w}\bar\cA_-}{3\kappa^{2}T^2} - \bar{\mathcal{A}}_{-} - \mathcal{A}_{+}  \right)
 \, ,
\\
B & \ = \ \frac{\partial_w \cA_+ \,  \partial_{\bar w} \cG - R \, \partial_{\bar w} \bar \cA_-   \partial_w \cG}{
R \, \partial_{\bar w}  \bar \cA_+ \partial_w \cG - \partial_w \cA_- \partial_{\bar w}  \cG} \,.
\label{eq:B}
\end{align}
Physically regular solutions corresponding to 5-brane junctions were constructed in~\cite{DHoker:2016ysh,DHoker:2017mds}, and solutions including additional 7-branes in~\cite{DHoker:2017zwj,Chaney:2018gjc}. They are specified in terms of the locally holomorphic functions $\cA_\pm$ whose detailed form can be found in these references.

\subsection{\texorpdfstring{${\rm AdS}_2\times\Sigma_{\fg_1}\times\Sigma_{\fg_2}\times S^2\times\Sigma$}{AdS2} solutions}

For each choice of the Riemann surface $\Sigma$ and locally holomorphic functions $\cA_\pm$ on $\Sigma$, the consistent truncations of~\cite{Hong:2018amk,Malek:2018zcz} provide a distinct uplift of solutions of six-dimensional $F(4)$ supergravity to solutions of type IIB supergravity. The effective six-dimensional Newton constant is related to the ten-dimensional Newton constant and the data specifying the Type IIB solution by~\cite{Gutperle:2018wuk,Fluder:2018chf}
\begin{align}\label{eq:6dNewton}
 \frac{1}{\kappa_6^2}& \ = \ \frac{32\pi}{3\kappa_{10}^2}\int \diff^2w \, \kappa^2\cG \, ,
\end{align}
where $2\kappa_{10}^2=(2\pi)^7{\alpha^\prime}^4$.

In particular, the ${\rm AdS}_2\times\Sigma_{\fg_1}\times\Sigma_{\fg_2}$ solutions of~\cite{Suh:2018tul}, which in addition to the metric involve a non-trivial scalar, $\SU(2)$ gauge field and two-form field, can be uplifted to Type IIB.
The resulting geometry takes the form
\begin{align}
 {\rm AdS}_2\times\Sigma_{\fg_1}\times\Sigma_{\fg_2}\times S^2 \times \Sigma \, , \qquad (\fg_1 > 1 \text{ and } \fg_2 > 1) \, ,
\end{align}
where ${\rm AdS}_2\times\Sigma_{\fg_1}\times\Sigma_{\fg_2}$ and $S^2$ are warped over $\Sigma$, with warp factors $\hat f_6^2$ and $\hat f_2^2$, respectively.
The type IIB supergravity fields resulting from the uplift (denoted by a hat on the warp factors) depend on all fields of the six-dimensional gauged supergravity, and agree with the solution presented in section~\ref{sec:ads6} only for the AdS$_6$ `vacuum' solution.
For the AdS$_2$ solutions of~\cite{Suh:2018tul} they can be constructed straightforwardly with the formulae of \cite{Hong:2018amk,Malek:2018zcz}. 

The advantage of having the consistent truncation is that many computations in six dimensions permit a clear ten-dimensional interpretation. This applies in particular to the relation between the on-shell action and the Bekenstein-Hawking entropy of magnetically charged AdS$_6$ black holes in the six-dimensional $F(4)$ supergravity derived in~\cite{Suh:2018tul}. Using that the on-shell action computes the sphere partition function of the dual field theory, the relation states 
\begin{align}\label{eq:SBH-FS5}
 S_{\rm BH}& \ = \ -\frac{8}{9}(1 - \fg_1 )( 1- \fg_2 )F_{S^5} \, .
\end{align}
With the consistent Kaluza-Klein reduction, this relation is implied to also hold for the uplifted ten-dimensional solutions in Type IIB. In ten dimensions, both computations involve an integral over the internal space, which reproduces the same factor that enters the computation of the effective six-dimensional Newton constant via \eqref{eq:6dNewton}\footnote{The sphere partition function can be extracted from the disc entanglement entropy \cite{Gutperle:2017tjo}, which involves an integral of $\hat f_6^4\hat f_2^2\hat\rho^2$ over $\Sigma$. The ten-dimensional horizon area involves an integral of the same combination of metric factors, which generically reduces to $\frac{2}{3}\kappa^2\cG$ for solutions uplifted via \cite{Hong:2018amk}.} and explains the universal relation from the Type IIB perspective.
This is analogous to the relation between sphere partition function and conformal central charge $C_T$ discussed in~\cite{Fluder:2018chf}.

\subsection{Solutions without monodromy: \texorpdfstring{$T_N$ and $\#_{N,M}$}{T-N and \#-N,M}}

The holomorphic functions and sphere partition functions for the $T_N$ and $\#_{N,M}$ solutions have been discussed in detail in~\cite{Fluder:2018chf}. The results for the sphere partition functions are
\begin{align}
 F_{{\rm sugra}}^{T_N}& \ = \  -\frac{27}{8\pi^2}\,\zeta(3)N^4\,,
 \qquad
 F_{{\rm sugra}}^{\#_{N,M}}   \ = \   -\frac{189}{16\pi^2}\,\zeta(3)N^2M^2\,.
\end{align}
With \eqref{eq:SBH-FS5} the black hole entropies thus become
\bea
 S_{\rm BH}^{T_N}& \ = \  \frac{3}{\pi^2}(1 - \fg_1)(1 - \fg_2)\,\zeta(3)N^4\,,
 \\
 S_{\rm BH}^{\#_{N,M}}&  \ = \   \frac{21}{2\pi^2}(1 - \fg_1)(1 - \fg_2)\,\zeta(3)N^2M^2\,.
\eea
In particular, they exhibit the same quartic scaling under an overall rescaling of the 5-brane charges in the brane junction defining the SCFT.

\subsection{Partition function and black hole entropy for \texorpdfstring{$T_{N,K,j}$}{T-NKj}}

The supergravity solution for the $T_{N,K,j}$ junction in general is a three-pole solution with one puncture with D7-brane monodromy.
For $N=jK$ it reduces to a ``minimal'' solution with only two 5-brane poles.

\subsubsection{The \texorpdfstring{$T_{N,K,j}$}{T-NKj} solution}
The locally holomorphic functions $\cA_\pm$ defining solutions with D7-branes take the general form
\begin{align}\label{eqn:cA-monodromy}
 \cA_\pm \ &= \  \cA_\pm^s + \cI \, , & \cI& \ = \ \int_\infty^w \diff z \;f(z)\sum_{\ell=1}^L \frac{Y^\ell}{z-r_\ell} \, ,
\end{align}
where $Y^\ell=Z_+^\ell-Z_-^\ell$ and $\cA_\pm^s$ correspond to a solution without monodromy,
\begin{align}\label{eqn:cA-0}
 \cA^s_\pm  & \ = \ \cA_\pm^0+\sum_{\ell=1}^L Z_\pm^\ell \log(w-r_\ell) \,, &
 Z_-^\ell \ &= \  - \overline{Z_+^\ell} \, .
\end{align}
The function $f$ encodes the branch cut structure,
\begin{align}
 f(w) & \ = \  \sum _{i=1}^I \frac{n_i^2}{4\pi} \log \left ( \gamma_i\,\frac{ w-w_i}{w -\bar w_i} \right ) .
\end{align}
The integration contour in \eqref{eqn:cA-monodromy} has to be chosen in such a way that no branch cuts are crossed.
The $T_{N,K,j}$ solution has three poles, $L=3$, and one puncture, $I=1$, and is realized by~\cite{Chaney:2018gjc}
\bea
 r_1& \ = \ 1 \, , & r_2& \ = \ 0 \, , & r_3& \ = \ -1 \, ,\\
  Z_+^1& \ = \ \frac{3}{8}\alpha^\prime \left(2N-\ii jK\right) \, ,
\quad &
 Z_+^2& \ = \ \frac{3}{4} \ii \alpha^\prime \left(N-jK\right) \, ,
\quad &
 Z_+^3& \ = \ -Z_+^1-Z_+^2 \, ,
\eea
with
\begin{align}
 n_1^2 \ &= \ j \,, & \gamma_1 \ &= \ 1 \,, &  w_1 \ &= \ \ii\tan\frac{\pi K}{2N} \, ,
\end{align}
and
\begin{align}
 \cA^0_+ \ &= \ Z_+^2\log 2+\frac{1}{2}\sum_{\ell=1}^L Y^\ell\int_\infty^{1} \diff x f^\prime(x)  \log |x-r_\ell|^2 \, .
\end{align}
As was shown in~\cite{Chaney:2018gjc}, this solves the regularity conditions, and stringy operators match between supergravity and field theory.

\subsubsection{Sphere partition function}

For the evaluation of the partition function it is convenient to obtain more explicit forms for $\cA_\pm$. That is, perform the polylogarithm integrals in \eqref{eqn:cA-monodromy}.
A crucial point is that the branch cut structure of the primitives has to be compatible with the contour chosen in \eqref{eqn:cA-monodromy}.

With the integration contour such that no branch cuts are crossed, we define
\begin{align}
 \mathfrak{f}(w,\alpha,r)& \ \equiv \ \int_{+\infty}^w \diff z\log\left(\frac{z-\ii\alpha}{z+\ii\alpha}\right)\frac{1}{z-r} \, ,
\end{align}
where $\alpha = \tan\frac{\pi K}{2N}$. An explicit expression with only the desired branch cut in the upper half-plane reads
\bea
 \mathfrak{f}(w,\alpha,r) \ = \ &
 \text{Li}_2\left(\frac{(w-\ii\alpha )(r+\ii \alpha )}{(w+\ii \alpha ) (r-\ii\alpha)}\right)
 +\text{Li}_2\left(\frac{2 \ii \alpha }{w+\ii \alpha }\right)
 -\text{Li}_2\left(\frac{r+\ii \alpha }{r-\ii \alpha }\right) \\ &
 +\log \left(\frac{w-\ii \alpha }{w+\ii \alpha }\right) 
 \log\left(\frac{2\ii \alpha  (r-w)}{(r-\ii \alpha ) (w+\ii\alpha)}\right) \, .
\eea
The locally holomorphic functions and their differentials are given by
\begin{align}
 \cA_\pm & \ = \   \cA_\pm^s + \cI \, , & \cI & \ = \   \frac{j}{4\pi}\sum_{\ell=1}^L Y^\ell \mathfrak{f}(w,\alpha,r_\ell) \, .
\end{align}
The expression for $\partial_w\cG$ becomes
\bea
 \partial_w\cG \ = \ &
 \left(\overline{\cA}{}_+^s-\cA_-^s\right)\partial_w\cA^s_+
 +\left(\cA_+^s-\overline{\cA}{}_-^s\right)\partial_w\cA^s_- \\ &
 +\left(\overline{\cA}{}_+^s-\cA_-^s+\cA_+^s-\overline{\cA}{}_-^s\right)\partial_w\cI
 +\left(\overline{\cI}-\cI\right)\left(\partial_w\cA^s_+-\partial_w\cA^s_-\right) .
\eea
The sphere partition function can be extracted from the finite part of the disc entanglement entropy, given by~\cite{Gutperle:2017tjo}
\begin{align}
 S_{\rm EE}^{\rm finite}& \ = \ -\frac{32\pi^3}{9 G_{\text{N}}}\int_\Sigma d^2w|\partial_w\cG|^2 \, ,
\end{align}
with $16\pi G_{\text{N}}=(2\pi)^7(\alpha^\prime)^4$.

\subsubsection{Explicit evaluation of partition function and entropy}

\begin{figure}
 \centering
 \includegraphics[width=0.45\linewidth]{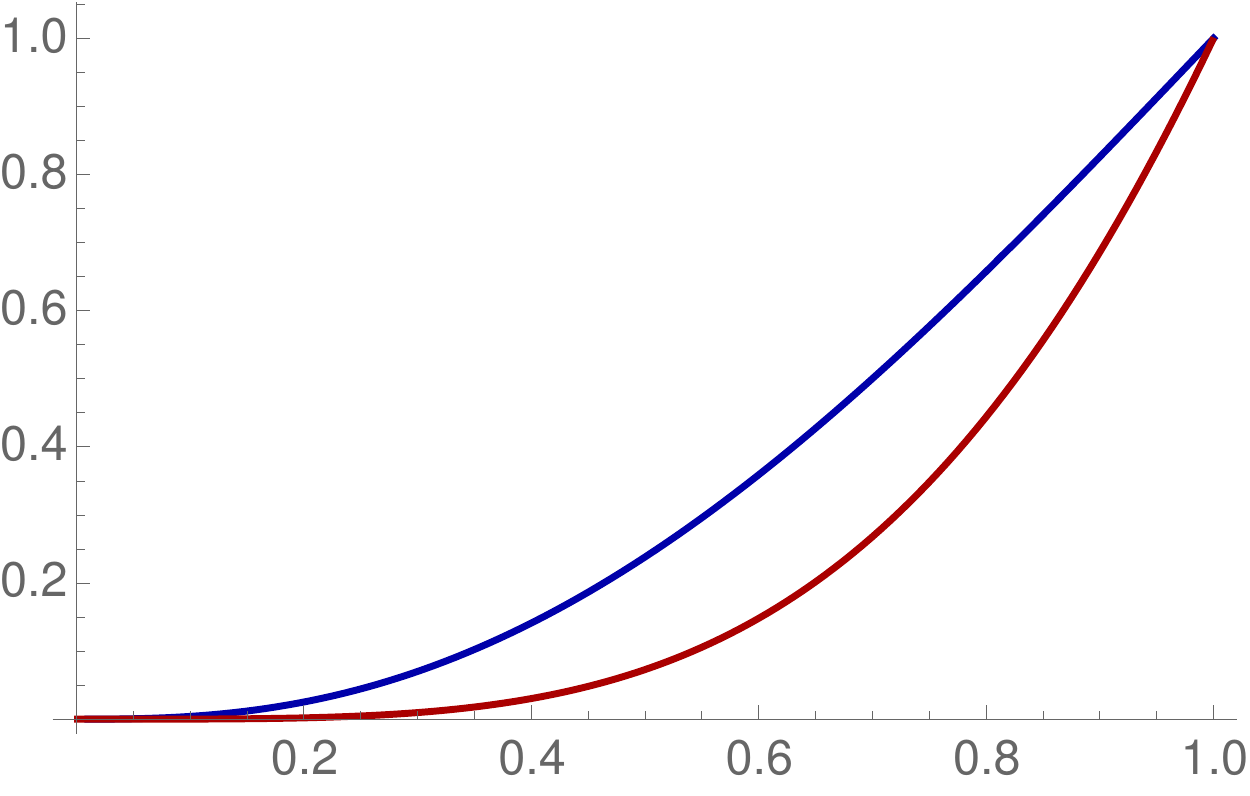}
 \put (5,5) {\small $k$}
 \caption{Plots of $\cS^{(1)}$ (upper curve) and $\cS^{(2)}$ (lower curve) as functions of $k\equiv K/N$. \label{fig:cS-1-2}}
\end{figure}
Under a simultaneous rescaling of $N$ and $K$ the functions $\cA_\pm$ scale linearly. As a result, the entanglement entropy scales quartically. This leaves two independent parameters,
\begin{align}
  S_{\rm EE}^{\rm finite}(N,K,j) \ &= \ N^4 \cS(k,j) \,, & k \ &\equiv \ \frac{K}{N} \, .
\end{align}
The functions $\cA_\pm$ are linear in $j$. The part independent of $j$ coincides with the holomorphic functions for the unconstrained $T_N$ theory, which we denote by $\cA_\pm^{T_N}$. The remaining part linear in $j$ is identical for $\cA_+$ and $\cA_-$, and we denote it by $\cX$, such that 
\begin{align}
 \cA_\pm \ = \ \cA_\pm^{T_N} + j\cX\,, 
 \qquad
 \cX \ = \ \partial_j(\cA_\pm^{s}+\cI)\,.
\end{align}
As a result of this decomposition, $\partial_w\cG$ is linear in $j$ as well, 
\bea
 \partial_w\cG \ & \ = \  \ \partial_w\cG^{T_N}+j\cZ \, ,
 \\
 \cZ \ & \ = \  \ 2 \, \mathrm{Re}\left(\cA_+^{T_N}-\cA_-^{T_N}\right)\partial_w\cX
 -2 \ii \, \mathrm{Im} (\cX)\left(\partial_w\cA_+^{T_N}-\partial_w\cA_-^{T_N}\right)\,.
\eea
This in turn implies that $\cS$ is a polynomial of degree two in $j$, which we parametrize as
\begin{align}
  \cS(k,j)& \ = \ \cS^{(0)}(k)\left[1-2j\cS^{(1)}(k)+j^2\cS^{(2)}(k)\right] \, .
\end{align}
The explicit expression for the finite part of the disc entanglement entropy  becomes
\begin{align}
 S_{\rm EE}^{\rm finite}& \ = \ -\frac{32\pi^3}{9 G_{\text{N}}}\int_\Sigma \diff^2w \Big[\left|\partial_w\cG^{T_N}\right|^2+2j \mathrm{Re}\left(\bar \cZ\partial_w\cG^{T_N}\right)+j^2\left|\cZ\right|^2\Big] \, .
\end{align}
Here, $\cS^{(0)}$ corresponds to the partition function of the unconstrained $T_N$ theory, \ie~
\begin{align}
  \cS^{(0)}& \ = \ -\frac{27}{8\pi^2} \zeta(3) \, .
\end{align}
The functions $\cS^{(1)}$ and $\cS^{(2)}$ can be determined numerically. Both vanish for $k=0$ and approach one for $k \to 1$.
Plots are shown in figure~\ref{fig:cS-1-2}. 
The finite part of the disc entanglement entropy thus can be written as
\begin{align}
 S_{\rm EE}^{\rm finite}(N,K,j)& \ = \ -\frac{27}{8\pi^2} \zeta(3)N^4\left[1-2j\cS^{(1)}(k)+j^2\cS^{(2)}(k)\right] \, .
\end{align}
Similarly, the black hole entropy, obtained using \eqref{eq:SBH-FS5}, is then given by
\begin{align}
 S_{\rm BH}(T_{N,K,j})& \ = \ \frac{3}{\pi^2}(1-\fg_1)(1-\fg_2)\,\zeta(3)N^4\left[1-2j\cS^{(1)}(k)+j^2\cS^{(2)}(k)\right] \, .
\end{align}

An interesting special case is when $N=2K$ and $j=2$, which contains the rank-$1$ $E_7$ theory for $K=2$ and provides a natural large $N$ generalization. As a further interesting example we take $j=1$ and $N = 2 K$, which realizes the $\chi_N^{N/2-1}$ theories of~\cite{Bergman:2014kza}. As a special case where $N>jK$, such that there is a non-vanishing number of unconstrained D5-branes, we include $j=3$ and $N=4K$. For these examples, we explicitly find
\bea\label{eq:FS5-sugra-TNKj}
S_{\rm EE}^{\rm finite}(T_{N,N/2,1})& \ = \ -0.245214\, N^4 \, , \\
S_{\rm EE}^{\rm finite}(T_{N,N/2,2})& \ = \ -0.139218\, N^4 \, ,\\
S_{\rm EE}^{\rm finite}(T_{N,N/4,3})& \ = \ -0.318856\, N^4 \, .
\eea
These values are smaller in absolute value than the analytic result for the unconstrained $T_N$ theory, consistent with a putative five-dimensional $F$-theorem. Likewise, one may flow from $T_{N,N/4,3}$ to $T_{N,N/2,2}$, and from $T_{N,N/2,1}$ to $T_{N,N/2,2}$.
The results are also consistent with a five-dimensional $F$-theorem for these cases.

\section{Discussion}\label{sec:discussion}

We have shown that the topologically twisted index of five-dimensional SCFTs that are defined on the intersection point of $(p,q)$ 5-brane junctions
computes the Bekenstein-Hawking entropy of a class of magnetically charged AdS black holes in Type IIB string theory.
The black hole entropy is obtained by uplifting the family of AdS$_2$ solutions to six-dimensional Romans' $F(4)$ gauged supergravity of~\cite{Suh:2018tul} to ten dimensions using the uplifts of~\cite{Hong:2018amk,Malek:2018zcz}.
The resulting ten-dimensional solutions have geometry AdS$_2 \times \Sigma_{\fg_1}\times\Sigma_{\fg_2}\times S^2\times\Sigma$, and are characterized by two locally holomorphic functions $\cA_\pm$ on the Riemann surface $\Sigma$. For each regular AdS$_6$ solution constructed in~\cite{DHoker:2016ujz,DHoker:2016ysh,DHoker:2017mds,DHoker:2017zwj,Chaney:2018gjc}, the uplift yields a regular ten-dimensional solution, which describes the near-horizon limit of a magnetically charged black hole with AdS$_6\times S^2 \times\Sigma$ asymptotics in Type IIB. In field theory terms, these AdS$_2$ solutions describe the twisted compactifications of the five-dimensional SCFTs dual to the associated AdS$_6$ solutions on a product of two Riemann surfaces. Our results show that the topologically twisted index agrees with the Bekenstein-Hawking entropy of the proposed dual black hole solution.
We have shown this for a representative sample of five-dimensional SCFTs, but the results confirm the general reasoning and are expected to extend to general pairs of supergravity solutions and five-dimensional SCFTs within the class of~\cite{DHoker:2016ujz,DHoker:2016ysh,DHoker:2017mds,DHoker:2017zwj,Chaney:2018gjc}. This in particular includes the solutions with 7-branes.

An important ingredient in evaluating the topologically twisted index, at large $N$, $M$, was the effective Seiberg-Witten prepotential $\cF(a)$ (that receives contributions from all the Kaluza-Klein modes on $S^1$)
of the dimensionally reduced four-dimensional theory living on $\Sigma_{\fg_1} \times \Sigma_{\fg_2}$.
It was conjectured in~\cite{Hosseini:2018uzp}, that the critical points of $\cF(a)$ combined with~\eqref{BAEs:general} dominate the large $N$ behavior of the twisted index.
In this paper, we employed this method and explicitly found agreement with the corresponding supergravity prescription, thus providing more evidence for the conjecture. It would be interesting to understand this from first principles, which we leave for future investigation.

A noteworthy observation in a similar spirit concerns a seemingly universal relation between the Seiberg-Witten prepotential and the twisted superpotential. In the case of Seiberg theories, it was (experimentally) found in~\cite{Hosseini:2018uzp}, that in the large $N$ limit the Seiberg-Witten prepotential $\cF$ and the twisted superpotential $\widetilde{\cW}$ satisfy
\be
\cF  \ = \ - \frac{2\pi \ii}{27} F_{S^{5}} \,, \qquad \widetilde{\cW} \ = \ \frac{4\pi \ii}{9} \left( 1- \fg_{2} \right) F_{S^{5}} \,.
\ee
For the theories in consideration in this paper, the numerics confirm the same relation,
suggesting that it might be universal at large $N$. It would be interesting to understand the physical reason behind this (including from a supergravity perspective).
In the case of the three-dimensional twisted index, the analogous relations can be understood from an insertion of a ``fibering operator"~\cite{Closset:2017zgf,Closset:2018ghr}. In the same vein, it would be interesting to derive a more unified framework for studying five-dimensional partition functions by the inclusion of similar ``geometry changing operators".

A perhaps curious observation in the localization computations is that the ratio of five-sphere partition function and topologically twisted index is captured accurately by the large $N$, $M$ asymptotics significantly earlier than the individual quantities.
For both quantities we have used large $N$, $M$ approximations, such as dropping instanton contributions, performing saddle point approximantions, and using asymptotic expansions of polylogarithms, in setting up the matrix models. Thus, neither quantity includes the full $N$, $M$ dependence.
Nevertheless, it is interesting to note that apparently the remaining leading $1/N$, $1/M$ corrections to the sphere partition function and the topologically twisted index cancel in the ratio.

Ultimately, it would be desirable to analytically understand and solve the matrix models computing the sphere partition functions and the topologically twisted indices.
It is interesting to note in that context, that we found the fundamental hypermultiplets in the quiver gauge theories discussed in section~\ref{sec:quivers} to be crucial for the matching of the leading large $N$, $M$ behavior between supergravity and field theory,
despite their naively subleading scaling compared to bi-fundamental and adjoint fields.
The universal relation between the topologically twisted index and the sphere partition function may also allow to combine analytic insights obtained from the corresponding matrix models to understand the large $N$, $M$ behavior analytically. 
A further direction for future investigation is to use potential consistent truncations to six-dimensional $F(4)$ gauged supergravity coupled to additional vector multiplets along the lines of~\cite{Malek:2019ucd} to study non-minimal twists.

\section*{Acknowledgements}
We thank Alberto Zaffaroni for useful discussions and comments.
This work used computational and storage services associated with the Hoffman2 Shared Cluster provided by UCLA Institute for Digital Research and Education's Research Technology Group.
MF is partially supported by the JSPS Grant-In-Aid for Scientific Research Wakate(A) 17H04837, and the WPI Initiative, MEXT, Japan at IPMU, the University of Tokyo.
The work of SMH was supported by World Premier International Research Center Initiative (WPI Initiative), MEXT, Japan.
The work of CFU is supported in part by the National Science Foundation under grant PHY-16-19926.

\begin{appendix}

\section{Matrix models for the \texorpdfstring{$\#_{N,M}$}{\#[N,M]}, \texorpdfstring{$T_{N}$}{T[N]} and \texorpdfstring{$T_{N,K,j}$}{T[N,K,j]} theories}\label{app:matrix-models}

In this appendix we present the relevant asymptotic expressions of the topologically twisted index and the corresponding matrix models for $T_N$, $\#_{N,M}$, and $T_{N,K,j}$ quiver gauge theories, which we discussed in section~\ref{sec:quivers}.
We are interested in the universal twist~\cite{Benini:2015bwz}, \ie\; 
\be
 \label{universal:twist:appendix}
 \Delta  \ = \  \pi \, , \qquad \fs  \ = \  1 - \fg_1 , \, \qquad \ft  \ = \  1 - \fg_2 \, .
\ee
We find it convenient to redefine the Coulomb branch parameters
\be
 a_\ell  \ = \  \ii t_\ell \, , \qquad \ell  \ = \  1 , \ldots , \text{rk}(G) \, ,
\ee
since they are purely imaginary (as confirmed by the numerical analysis). In the large $N$, $M$ limit $t_\ell$ grows with some positive powers of $N$ and $M$, and therefore we can approximate the polylogarithms appearing in the matrix models
(see~\eqref{index:Sigmag1xSigmag2xS1},~\eqref{pert:twisted:superpotential} and~\eqref{pert:prepotential}) by their asymptotic forms given in~\eqref{PolyLog:asymptotic}. Furthermore, in such a limit, instanton contributions to the topologically twisted index are exponentially suppressed. Hence, we only need to consider the perturbative contributions to the index.

Let us start with the effective Seiberg-Witten prepotential $\cF(a)$. The contributions of a vector multiplet and a hypermultiplet to~\eqref{pert:prepotential}, for $t \to \infty$, read%
\footnote{We set the circumference of $S^1$ to one ($\beta = 1$) throughout this section.}
\bea\label{Fasymp}
 2 \pi \ii \cF_{\text{V}} (t) &  \ = \  - \frac{\ii}{2} g_3 ( - \ii t ) \sign ( t ) + \zeta (3) + \ldots \, , \\
 2 \pi \ii \cF_{\text{H}} (t) &  \ = \  - \frac{\ii}{2} g_3 ( \ii t + \pi ) \sign ( t ) - \zeta (3) + \ldots \, ,
\eea
respectively, where the ellipses denote subleading terms. 

Next, we compute the (asymptotic) contributions of a vector multiplet and a hypermultiplet to the effective twisted superpotential~\eqref{pert:twisted:superpotential}. They can be written as
\bea\label{Wasymp}
 \wt\cW_{\text{V}} (t , \fn) &  \ = \  - \frac{1}{2} ( \fn + 1 - \fg_2 ) g_2 ( - \ii t ) \sign ( t ) + \ldots \, , \\
 \wt \cW_{\text{H}} (t , \fn) &  \ = \  \frac{1}{2} \fn \, g_2 ( \ii t  + \pi ) \sign ( t ) + \ldots \, ,
\eea
respectively, where the ellipses again denote the subleading terms in the $t\to \infty$ limit. 

Finally, in the case of the universal twist~\eqref{universal:twist:appendix} only vector multiplets contribute to the topologically twisted index in~\eqref{Bethe:sum:Z} directly. The hypermultiplet contributions to the index only enter through the Bethe ansatz equations~\eqref{BAEs:general} and~\eqref{MAEs:general}. At large $t$ (or equivalently $N$, $M$), we obtain the following expression for the logarithm of the twisted index
\bea\label{logZasymp}
 \log Z_{\text{V}} (t , \fn) &  \ = \  \frac{\ii}{2} (1 - \fg_1 ) ( \fn + 1 - \fg_2 ) g_1 ( - \ii t ) \sign ( t ) + \ldots \, ,
\eea
where again the ellipses denote the subleading terms in the $t\to \infty$ limit.

\subsection{\texorpdfstring{$T_N$}{T-N} theories}

We start with the $T_N$ theories, whose infrared gauge theory description is given in~\eqref{TNquiver}. The effective Seiberg-Witten prepotential is given by
\bea
 \cF_{T_N} ( t )  \ = \ & 
 \sum_{\substack{\ell , m = 1 \\ \ell \neq m}}^{j} \sum_{j = 2}^{N-1} \cF_{\text{V}} \big( t^{(j)}_{\ell} - t^{(j)}_{m} \big)
 + \sum_{\ell = 1}^{j} \sum_{m = 1}^{j + 1} \sum_{j = 2}^{N-2} \cF_{\text{H}} \big( t^{(j)}_{\ell} - t^{(j + 1)}_{m} \big) \\ &
 + 2 \sum_{\ell = 1}^{2} \cF_{\text{H}} \big( t^{(2)}_{\ell} \big) + N \sum_{\ell = 1}^{N-1} \cF_{\text{H}} \big( t^{(N-1)}_{\ell} \big) \,,
\eea
where we additionally impose the constraint 
\be\label{TN:sucondt}
\sum_{\ell =1}^{j} t_{\ell}^{(j)} \ = \ 0 \, , \qquad  j = 2 , \ldots , N-1 \, .
\ee
Similarly, the effective twisted superpotential reads
\bea
 \wt \cW_{T_N} ( t , \fn )  \ = \ & \sum_{\substack{\ell , m = 1 \\ \ell \neq m}}^{j} \sum_{j = 2}^{N-1} \wt \cW_{\text{V}} \big( t^{(j)}_{\ell} - t^{(j)}_{m} , \fn^{(j)}_{\ell} - \fn^{(j)}_{m} \big) \\ &
 + \sum_{\ell = 1}^{j} \sum_{m = 1}^{j + 1} \sum_{j = 2}^{N-2} \wt \cW_{\text{H}} \big( t^{(j)}_{\ell} - t^{(j + 1)}_{m} , \fn^{(j)}_{\ell} - \fn^{(j + 1)}_{m} \big) \\ &
 + 2 \sum_{\ell = 1}^{2} \wt \cW_{\text{H}} \big( t^{(2)}_{\ell} , \fn^{(2)}_{\ell} \big) + N \sum_{\ell = 1}^{N-1} \wt \cW_{\text{H}} \big( t^{(N-1)}_{\ell} , \fn^{(N-1)}_{\ell} \big) \,,
\eea
where in addition to~\eqref{TN:sucondt} we further require
\be\label{TN:sucondn}
\sum_{\ell =1}^{j} \fn_{\ell}^{(j)} \ = \ 0 \, , \qquad  j = 2 , \ldots , N-1 \, .
\ee
Finally, the topologically twisted index can be written as
\be
 \log Z_{T_N} (t , \fn)  \ = \  \sum_{\substack{\ell , m = 1 \\ \ell \neq m}}^{j} \sum_{j = 2}^{N-1}
 \log Z_{\text{V}} \big( t_\ell^{(j)} - t_m^{(j)} , \fn^{(j)}_{\ell} - \fn^{(j)}_{m}  \big) \, .
\ee

\subsection{\texorpdfstring{$\#_{N,M}$}{\#-N,M} theories}

Now, we turn to the $\#_{N,M}$ theories, whose gauge theory quiver description is given in~\eqref{eq:D5NS5-quiver}. The effective Seiberg-Witten prepotential reads
\bea
 \cF_{\#_{N,M}}( t )  \ = \ & \sum_{j = 1}^{M-1} \sum_{\substack{\ell , m = 1 \\ \ell \neq m}}^{N} \cF_{\text{V}} \big( t^{(j)}_{\ell} - t^{(j)}_{m} \big)
 + \sum_{j = 1}^{M-2} \sum_{\ell , m = 1}^{N} \cF_{\text{H}} \big( t^{(j)}_{\ell} - t^{(j + 1)}_{m} \big) \\ &
 + N \sum_{\ell = 1}^{N} \left[ \cF_{\text{H}} \big( t^{(1)}_{\ell} \big) + \cF_{\text{H}}\big( t^{(M-1)}_{\ell} \big) \right] \,,
\eea
where the eigenvalues obey the constraint
\be
 \sum_{\ell=1}^{N} t_\ell^{(j)}\ = \ 0 \, , \qquad j = 1 , \ldots, M-1 \, .
\ee
The effective twisted superpotential can be written as
\bea
 \wt \cW_{\#_{N,M}} ( t , \fn ) \ = \  & 
 \sum_{j = 1}^{M-1} \sum_{\substack{\ell , m = 1 \\ \ell \neq m}}^{N} \wt \cW_{\text{V}} \big( t^{(j)}_{\ell} - t^{(j)}_{m} , \fn^{(j)}_{\ell} - \fn^{(j)}_{m} \big) \\ 
 & + \sum_{j = 1}^{M-2} \sum_{\ell , m = 1}^{N} \wt \cW_{\text{H}} \big( t^{(j)}_{\ell} - t^{(j + 1)}_{m} , \fn^{(j)}_{\ell} - \fn^{(j + 1)}_{m} \big) \\ 
 & + N \sum_{\ell = 1}^{N} \left[ \wt \cW_{\text{H}} \big( t^{(1)}_{\ell} \big) + \wt \cW_{\text{H}}\big( t^{(M-1)}_{\ell} \big) \right] \,,
\eea
with the constraint on the gauge magnetic fluxes
\be
 \sum_{\ell=1}^{N} \fn_\ell^{(j)}\ = \ 0 \, , \qquad j = 1 , \ldots, M-1 \, .
\ee
Finally, the topologically twisted index is given by
\be
 \log Z_{\#_{N,M}} (t , \fn)  \ = \  
 \sum_{j = 1}^{M-1} \sum_{\substack{\ell , m = 1 \\ \ell \neq m}}^{N} \log Z_{\text{V}} \left( t_\ell^{(j)} - t_m^{(j)} , \fn^{(j)}_{\ell} - \fn^{(j)}_{m} \right) \, .
\ee

\subsection{\texorpdfstring{$T_{N,K,j}$}{T-N,K,j} theories}

Finally, let us consider the $T_{N,K,j}$ theories, whose gauge theory descriptions are given in the quivers~\eqref{eq:TNKj-quiver-1},~\eqref{eq:TNKj-quiver-2}, and~\eqref{eq:TNKj-quiver-3}. We explicitly spell out the example with $j=2$ and $N=2K$, \emph{i.e.}\;$T_{2K,K,2}$, with $K > 2$. The corresponding effective Seiberg-Witten prepotential reads
\bea
 \label{Seiberg-Witten:F:j=2:N=2K}
 \cF_{T_{2K,K,2}} ( t )  \ = \ & 
 \sum_{\substack{\ell , m = 1 \\ \ell \neq m}}^{r} \sum_{r = 2}^{K} \cF_{\text{V}} \big( t^{(r,R)}_{\ell} - t^{(r,R)}_{m} \big)
 + \sum_{\ell = 1}^{r} \sum_{m = 1}^{r + 1} \sum_{r = 2}^{K - 1} \cF_{\text{H}} \big( t^{(r,R)}_{\ell} - t^{(r + 1,R)}_{m} \big) \\ &
 + \sum_{\substack{\ell , m = 1 \\ \ell \neq m}}^{r} \sum_{r = 2}^{K - 1} \cF_{\text{V}} \big( t^{(r,L)}_{\ell} - t^{(r,L)}_{m} \big)
 + \sum_{\ell = 1}^{r} \sum_{m = 1}^{r + 1} \sum_{r = 2}^{K - 2} \cF_{\text{H}} \big( t^{(r,L)}_{\ell} - t^{(r + 1,L)}_{m} \big) \\ &
 + 2 \sum_{\ell = 1}^{2} \cF_{\text{H}} \big( t^{(2,R)}_{\ell} \big) 
 + 2 \sum_{\ell = 1}^{K} \cF_{\text{H}} \big( t^{(K,R)}_{\ell} \big) 
 + 2 \sum_{\ell = 1}^{2} \cF_{\text{H}} \big( t^{(2,L)}_{\ell} \big) \\ &
 + \sum_{\ell = 1}^{K-1} \sum_{m = 1}^{K} \cF_{\text{H}} \big( t^{(K-1,L)}_{\ell} - t^{(K,R)}_{m} \big) \, .
\eea
Here, we introduced the notation $t_{\ell}^{(r,L)}$, $t_{\ell}^{(r,R)}$ to label the Coulomb branch parameters on the left, right hand side of the central $\SU(K)$ gauge group in~\eqref{eq:TNKj-quiver-3}, respectively. The expression for the twisted superpotential $ \wt \cW_{T_{2K,K,2}} ( t , \fn )$ is very similar --- one needs to replace $\cF_{\text{V}}(t)$, $\cF_{\text{H}}(t)$ with $\wt \cW_{\text{V}}(t,\fn)$, $\wt \cW_{\text{H}}(t,\fn)$ in~\eqref{Seiberg-Witten:F:j=2:N=2K}, respectively. In addition, we have to impose the following conditions
\be
\sum_{\ell=1}^{r} t_{\ell}^{(r,L)} \ = \ 0 \,, \qquad \text{and} \qquad \sum_{\ell=1}^{r} t_{\ell}^{(r,R)} \ = \ 0 \,,
\ee
as well as analogous constraints on the gauge magnetic fluxes $\fn_{\ell}^{(r,L)}$ and $\fn_{\ell}^{(r,R)}$.
Lastly, the topologically twisted index can be written as
\bea
 \log Z_{T_{2K,K,2}} (t , \fn)  \ = \ & \sum_{\substack{\ell , m = 1 \\ \ell \neq m}}^{r} \sum_{r = 2}^{K} \log Z_{\text{V}} \big( t^{(r,R)}_{\ell} - t^{(r,R)}_{m} , \fn^{(r,R)}_{\ell} - \fn^{(r,R)}_{m} \big) \\ &
 + \sum_{\substack{\ell , m = 1 \\ \ell \neq m}}^{r} \sum_{r = 2}^{K - 1} \log Z_{\text{V}} \big( t^{(r,L)}_{\ell} - t^{(r,L)}_{m} , \fn^{(r,L)}_{\ell} - \fn^{(r,L)}_{m} \big) \, .
\eea

\section{Five-sphere partition function}

Let us briefly recall the necessary ingredients for the numerical computation of the five-sphere partition function for the $T_{N,k,j}$ theories. As mentioned in the main text, we expect that the partition function of the superconformal fixed point is reproduced in the infrared gauge theory, which relies on the assumption that the relevant higher-order derivative corrections are $\cQ$-exact, and thus not relevant for the partition function. Furthermore, general arguments in~\cite{Jafferis:2012iv} suggest that the nonperturbative (instanton) contributions to the partition function are suppressed in the large $N$ limit. We refer to~\cite{Jafferis:2012iv} for more details.\footnote{See also~\cite{Chang:2017cdx,Chang:2017mxc,Fluder:2018chf} for evidence that the instantonic contributions are ``small" compared with the perturbative piece even at small $N$.} Thus, for the purposes of this paper, we may solely look at the perturbative part of the five-sphere partition function. This was computed using supersymmetric localization~\cite{Pestun:2007rz} on the (round) five-sphere in~\cite{Kallen:2012va,Kim:2012ava} and for the squashed five-sphere in~\cite{Imamura:2012bm} (see also~\cite{Lockhart:2012vp} for the same result derived from topological strings). We shall use the latter reference, and set the squashing parameters to vanish.

The perturbative part $Z^{S^5}_{\rm pert}$ of the five-sphere partition function of a gauge theory with gauge group $G$ of rank $\text{rk}(G)$, with $I=1, \ldots, N_{f}$ hypermultiplets in a representation $\oplus_{I} (\fR_I \oplus \bar \fR_I) $, of the gauge group $G$ is given by
\bea\label{S5pert}
Z^{S^5}_{\rm pert} \ = \ \frac{ 2\pi e^{\frac{\zeta(3)}{4\pi^{3}}}}{\left| \mathfrak{W} \right|} \int_{-\infty}^{\infty}\prod_{\ell=1}^{\text{rk}(G)} \frac{\diff a_\ell}{2\pi} \, e^{- \mathfrak{F}(a)} \frac{\prod_{\alpha \in G} S_{3} \left( - \ii  \alpha(a) \mid 1,1,1 \right)}{\prod_{I=1}^{N_f} \prod_{\rho_I \in \fR_I} S_{3} \left( \ii \rho_{I} (a)+\frac{3}{2} \mid 1,1,1 \right)}\,,
\eea
where the products are over all the roots $\alpha$ of $G$ and weights $\rho_{I}$ of the relevant representation $\fR_{I}$, by $|\mathfrak{W}|$ we denote the cardinality of the Weyl group of $G$, and $\zeta(x)$ is the Riemann zeta function. Furthermore, $\mathfrak{F}(a)$ is proportional to the classical piece of the (flat space) Seiberg-Witten prepotential (see~\cite{Intriligator:1997pq}). For vanishing Chern-Simons contributions, $\mathfrak{F}(a)$ is in fact subleading, and thus not relevant for our purposes here. Lastly, $S_{3}(x \mid 1,1,1)$ is the triple-sine function with $\omega_i=1$, $i=1,2,3$. It can be defined as
\be
S_{3} \left( z \mid 1,1,1 \right) \ \equiv \ \exp\left( -\frac{\pi \ii}{6} B_{3,3}\left( z \mid 1,1,1 \right) - I_{3} \left( z \mid 1,1,1 \right)\right) \,,
\ee
where $ B_{3,3}\left( z \mid 1,1,1 \right)$ is the generalized Bernoulli polynomial given by
\be
 B_{3,3}\left( z \mid 1,1,1 \right) \ = \ z^{3} -\frac{9}{2} z^{2} + 6 z - \frac{9}{4} \, ,
\ee
and $ I_{3} \left( z \mid 1,1,1 \right)$ can be explicitly computed in terms of the following integral
\be
 I_{3} \left( z \mid 1,1,1 \right) \ = \ \int_{\mathbb{R}+ \mathrm{i} 0^{+}} \frac{\diff x}{x} \frac{e^{z x}}{\left( e^{x} -1 \right)^{3}} \, ,
\ee
where the contour runs over the real axis with a semi-circle around $x=0$ going into the positive half-plane. 

In this paper, we evaluate the five-sphere partition functions for $T_{N,K,j}$ theories numerically, to extract the large $N$ limit and compare to supergravity. The relevant matrix models can be generated in analogy with the ones for the twisted indices in appendix~\ref{app:matrix-models}, where we replace the vector multiplet and hypermultiplet contributions with the relevant pieces of the five-sphere partition function, \emph{i.e.}
\bea
\cF_{\text{V}}(t) \ \to \ & \cF^{S^5}_{\text{V}}(t) \ = \ -\frac{1}{2}\Big[  \log S_3 \left( \ii t \mid 1,1,1 \right) +\log S_3 \left( - \ii t \mid 1,1,1 \right)  \Big] \,, \\
\cF_{\text{H}}(t) \ \to \ & \cF^{S^5}_{\text{H}}(t) \ = \  \log S_3 \left( \ii t +\frac{3}{2} \mid 1,1,1 \right) \,,
\eea
and with additional overall contributions, which are not relevant in the large $N$ limit.

\end{appendix}

\bibliographystyle{ytphys}

\bibliography{BHTN}

\end{document}